# Representations and Strategies for Transferable Machine Learning Models in Chemical Discovery

Daniel R. Harper[1,2,#], Aditya Nandy[1,2,#], Naveen Arunachalam[1], Chenru Duan[1,2], Jon Paul Janet[1], and Heather J. Kulik[1,*]

[1]*Department of Chemical Engineering, Massachusetts Institute of Technology, Cambridge, MA 02139*

[2]*Department of Chemistry, Massachusetts Institute of Technology, Cambridge, MA 02139*

[#]These authors contributed equally.

ABSTRACT: Strategies for machine-learning(ML)-accelerated discovery that are general across materials composition spaces are essential, but demonstrations of ML have been primarily limited to narrow composition variations. By addressing the scarcity of data in promising regions of chemical space for challenging targets like open-shell transition-metal complexes, general representations and transferable ML models that leverage known relationships in existing data will accelerate discovery. Over a large set (ca. 1000) of isovalent transition-metal complexes, we quantify evident relationships for different properties (i.e., spin-splitting and ligand dissociation) between rows of the periodic table (i.e., $3d/4d$ metals and $2p/3p$ ligands). We demonstrate an extension to graph-based revised autocorrelation (RAC) representation (i.e., eRAC) that incorporates the effective nuclear charge alongside the nuclear charge heuristic that otherwise overestimates dissimilarity of isovalent complexes. To address the common challenge of discovery in a new space where data is limited, we introduce a transfer learning approach in which we seed models trained on a large amount of data from one row of the periodic table with a small number of data points from the additional row. We demonstrate the synergistic value of the eRACs alongside this transfer learning strategy to consistently improve model performance. Analysis of these models highlights how the approach succeeds by reordering the distances between complexes to be more consistent with the periodic table, a property we expect to be broadly useful for other materials domains.

# 1. Introduction.

Machine learning (ML) models have been demonstrated as a powerful alternative to conventional computational high-throughput screening[1-2] by greatly accelerating chemical discovery[3-8], e.g., with active learning.[9-12] It is typically the case at the outset of a discovery campaign that a great deal is known from experiments, computation, and associated ML models about a highly localized region of chemical space, but the most promising directions for improvement lie outside this region. For example, many of the most active catalysts[13-14] or single molecule magnets[15-16] require the use of rare and toxic $4d$ or $5d$ metals, and identification of principles[17] for their replacement with $3d$ metals would be advantageous[18-19] in a number of difficult small molecule catalytic conversions (e.g., $H_2$[20-25], $CO_2$[26], and $O_2$[27]). Thus, ML-accelerated discovery requires models and representations that generalize well and strategies to exploit existing knowledge while data in new regions is scarce.

Despite the need for ML models that generalize across the periodic table[3, 28-31], ML models have much more commonly been applied in narrow regions of chemical space, such as closed-shell organic (i.e., CHNOF-containing) molecules in the QM9 dataset[32], with few exceptions[33]. For open-shell transition-metal chemistry in particular, unique challenges are present in the generation of sufficiently large data sets for ML model training while spanning a large range of metal centers, ligand chemistry, and the cooperative effect the metal/ligand identity play in the resulting spin or oxidation state of the complex.[34] While ML-model-accelerated computational discovery has targeted numerous transition-metal catalysts[35-39] or functional complexes[9, 40] and related metal–organic materials (e.g., metal–organic frameworks[41-43] and transition metal oxides[44]), these models have also generally been restricted to a small number of transition metals or modest modifications of the ligands. While only preliminarily demonstrated on open-shell transition-metal complexes[45-47], there is evidence that electronic descriptors such as partial charges or bond valence[45] and molecular orbitals[46, 48-49] or other computed electronic properties[50-51] are one path to training transferable ML models on small data sets. Sidestepping ML with direct, computationally demanding simulation (e.g., with density functional theory or DFT) in large spaces of functional materials and catalysts[34, 52] is plagued by combinatorial challenges[34, 52-54] that can only be partly reduced through inverse design strategies (e.g., with alchemical derivatives[54-57] or heuristics[58]). Nevertheless, all such approaches that require an electronic structure calculation to be completed for property prediction or optimization hamper the large-scale exploration that is required for large-scale materials discovery.

For transferable ML models, opportunities and challenges for either conventional 3D-structure-based[59-63] or graph-based[64-66] representations remain. Here, we focus on graph-based revised autocorrelations[65, 67-68] (RACs[65]) that are no-cost products and differences of heuristic atomic properties (e.g., nuclear charge or Pauling electronegativity) on the molecular graph. The 155-feature form of RACs (i.e., RAC-155[65]) was used[9] to train a multi-task artificial neural network (ANN) that accelerated identification of soluble and high redox potential TMCs for redox flow batteries from nearly 3M candidates in weeks instead of decades.[9] In this or other discovery campaigns for which there is a reasonable degree of similarity between training and test data, RAC-155 and its feature-selected subsets have been used[9, 35, 38, 65, 69-72] to train ML (e.g., kernel ridge regression, KRR, and ANN) models that exhibit good performance on modestly sized (ca. 200–1000 data points) data sets of DFT properties for TMCs, including around 1–4 kcal/mol accuracy for adiabatic spin-splitting energies[65, 71], bond lengths[65, 69, 73], ionization/redox potentials[65], HOMO energies,[70] and catalytic properties.[38]

Exemplary of how RAC-trained models have limited domains of applicability, Janet et al.

observed[71] that introducing new elements (e.g., P or As) into an out-of-distribution test molecule uniquely challenged ANN models trained only on first-row, $3d$ open-shell transition-metal complexes with primarily $2p$ ligand chemistry. Conversely, introducing diverse metal coordination environments with a wider range of sampled elements into model training greatly improved ANN model generalization[74], reducing mean absolute errors by around 25–30%. Because distance in feature space (e.g., in KRR) or latent space (i.e., in the ANN) is a critical component, prediction by ML models of new materials with compositions not well supported by training data is universally challenging. Nevertheless, RACs and other widely used representations[60] employ the nuclear charge in the representation, exaggerating differences between rows of the periodic table with respect to differences in group number[75]. While some heuristic properties in RACs (e.g., the Pauling electronegativity) more faithfully encode chemical similarity[64], the nuclear-charge-based features are often emphasized in feature-selected subsets, highlighting their importance in model prediction.[69]

Limitations in the representation will make it challenging for an ML model to learn relationships between isovalent materials with elements from different rows of the periodic table, despite clear relationships[76-78] in both ligand field strength (e.g., $3p$ P > S and $2p$ N > O) and $d$-filling trends of spin-splitting energies[79] between $3d$ and $4d$ TMCs. One may anticipate that changing the row of the metal requires a change in the ligand chemical composition (e.g., for nitrogen-coordinating ligands vs phosphorus-coordinating ligands) as well. The existence of these structure-property relationships presents an opportunity for representation improvement. In this work, we tackle the challenge of improving the transferability of no-cost representations by introducing effective nuclear charge as a heuristic into RACs, an approach previously demonstrated only in closed-shell organic molecules[80-81] or in metallic alloys[33, 82]. We show how this always improves model performance and devise a synergistic strategy for leveraging large amounts of data containing elements from one row of the periodic table to improve the learning curves on materials belonging to a new target row. While demonstrated here on 3$d$/4$d$ and 2$p$/3$p$ learning tasks, we expect this approach to be broadly useful for other materials.

2. Approach.

2a. Dataset Construction.

Mononuclear, octahedral TMCs were assembled from combinations of eight transition metals with ten small, common ligands to explore relationships among isovalent metal centers and metal-coordinating ligand atoms. While the electronic structure of some of these TMCs were discussed in Ref. [79], we reiterate the rationale for their curation in paired datasets for ML on isovalent properties here. The TMCs contain 3$d$ or 4$d$ mid-row transition metals (Cr/Mo, Mn/Tc, Fe/Ru, and Co/Rh) in +2 and +3 oxidation states (Figure 1). We study the energetically accessible spin states for these metals, which correspond to low-spin (LS), intermediate-spin (IS), and high-spin (HS) states, where the IS and HS states differ by having two more or four more unpaired electrons, respectively, with respect to the singlet or doublet LS state. The earliest Cr(III)/Mo(III) or latest Co(II)/Rh(II) are studied in only two states, a LS doublet and an IS quartet (Supporting Information Table S1).

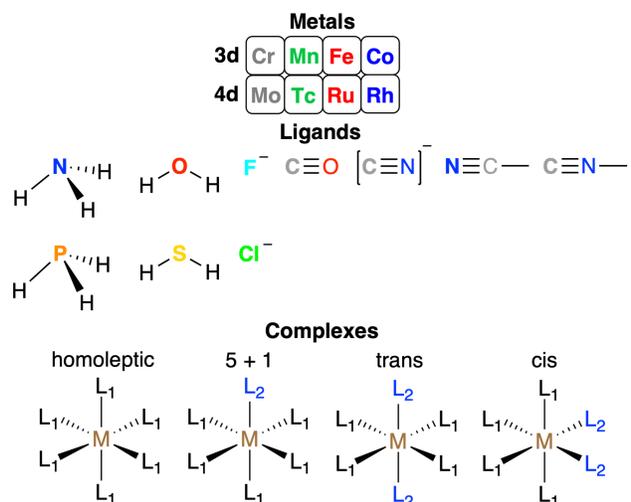

**Figure 1.** Overview of eight metals and ten ligands used to form mononuclear, octahedral transition-metal complexes in the data sets in this work. (top) 3$d$ and 4$d$ metals grouped by row and column order in the periodic table. (middle) Skeletal structures of the ten ligands with the metal-coordinating atom shown in bold. The six left-most ligands are grouped by isovalent 2$p$ metal-coordinating atoms in the first row (i.e., N, O, and F) with 3$p$ metal-coordinating atoms in the second row (P, S, and Cl). Four additional ligands that have carbon or nitrogen coordinating atoms are depicted at right. (bottom) Symmetries of transition-metal complexes studied by assembling up to two ligand types ($L_1$ in black and $L_2$ in blue) in octahedral complexes: homoleptic ($L_1 = L_2$), 5+1 (a single axial $L_2$), and trans or cis placement of two $L_2$ ligands.

The ten common ligands were selected for their varying field strengths (i.e., from weak field halides to strong field carbonyl). Six (i.e., three pairs) of these ligands contain isovalent coordinating atoms that differ in principal quantum number, e.g., 2$p$ vs. 3$p$ $NH_3$ and $PH_3$ (Figure 1). A large set of possible TMC isomers were assembled from up to two of these ligands (i.e., $L_1$ and $L_2$) in four arrangements: i) homoleptic $M(L_1)_6$; ii) trans $M(L_1)_4(L_2)_2$; iii) cis $M(L_1)_4(L_2)_2$; and iv) 5+1 $M(L_1)_5(L_2)$ (Figure 1). All homoleptic (i.e., 440) TMCs and trans or 5+1 (i.e., 3,960 each) structures were set up for calculation with DFT, whereas only a subset (1,080) of the cis structures were set up (see Sec. 3a and Supporting Information Table S2). For the TMCs in this set, we compute (see Sec. 3a) up to three gas-phase, adiabatic spin-splitting energies, i.e., between the LS and HS states, $\Delta E_{H-L}$, as well as between the IS state and either the LS or the HS state (i.e., $\Delta E_{H-I}$ and $\Delta E_{I-L}$), which were discussed in part in Ref. [79]. In this work, for a subset of the TMCs, we now also compute a vertical ligand dissociation (LD) energy, $\Delta E_{LD}$, the energy required to rigidly remove a single axial ligand (see Sec. 3a).

From nearly ten thousand hypothetical TMCs, we obtain a smaller subset for testing the transferability of ML models by pairing isovalent species with differing principal quantum numbers via a filtering procedure that is more restrictive than that used in Ref. [79]. For the isovalent metal pairing (IMP) set, we compare computed properties of pairs of complexes that have metal centers in the same group of the periodic table but differ in their principal quantum number (e.g., 3$d$ Fe(II) and 4$d$ Ru(II)) with identical characteristics for all other TMC properties (e.g., spin multiplicity, $L_1$/$L_2$ ligands, and symmetry class). After accounting for data fidelity and convergence rates, we obtain over 1200 pairs of spin-splitting $\Delta E$ values depending on the spin states compared, with the I-L set roughly twice as large as the H-L or H-I sets. The IMP set of $\Delta E_{LD}$ values across spin states contains over 1000 pairs (Supporting Information Table S3).

We also construct an isovalent ligand pairing (ILP) set, which compares properties of pairs of TMCs that have ligands with metal-coordinating atoms that belong to the same group of the periodic table but differ in principal quantum number (e.g., 2$p$ N and 3$p$ P). Because only six of the ten ligands belong to one of these 2$p$/3$p$ pairs, heteroleptic TMCs could include between zero and two unique ligands from this subset. For the pairs in the ILP set, we change all relevant ligands simultaneously while holding any incompatible ligands and all other TMC properties (e.g. metal, oxidation and spin state) fixed. Because we start from a smaller subset of TMCs that have the relevant ligands, ILP subsets of spin-splitting and ligand dissociation energies are about half of the size of the IMP sets (Supporting Information Table S4).

**2b. Representations.**

We employ Moreau-Broto autocorrelations (ACs)[67, 83] and revised AC variants (i.e.,

RACs)[65], which are representations formed by discrete operations on the heuristic properties of atoms in the molecular graph in analogy to continuous autocorrelations. A standard AC is evaluated over all of the *n* atoms in a molecule that are *d* bonds apart:

$$P_d = \sum_i^{start} \sum_j^{scope} P_i P_j \delta(d_{ij}, d) \qquad (1)$$

where $P$ is the chosen property, $P_i$ and $P_j$ are the specific values of the *i*th and *j*th atoms that are $d_{ij}$ bonds apart. The five heuristic properties[65, 67-68, 83-84] we have employed for transition-metal chemistry[65, 84] include: the Pauling electronegativity, nuclear charge, covalent radius, topology, and the identity (i.e., 1) and produce a $5d+5$ dimensional vector. ACs have achieved best-in-class performance for geometry-free descriptor predictions of atomization energies on subsets of QM9[32] using KRR[65] or ANN models[71] with diminishing returns observed for depth cutoffs above three[65], although higher depth cutoffs are sometimes used[68, 84].

Extensions of ACs to enrich metal-local descriptors[85] in the feature vector led to the development of RACs, as first introduced in Ref. [65]. RACs include centering (i.e., always including in the sum) the property evaluation on the metal (i.e., mc-RACs) or the ligand atom that coordinates the metal (i.e., lc-RACs). The lc-RACs are averaged separately over ligands of specific types, which were referred to as the scope in Ref. [65]. In the octahedral complexes studied here, lc-RACs are averaged separately over the equatorial or axial ligands. RACs also introduced the difference of two atomic properties:

$$P'_d = \sum_i^{start} \sum_j^{scope} (P_i - P_j) \delta(d_{ij}, d) \qquad (2)$$

which can be non-trivially computed for metal-centered or ligand-centered RACs excluding the identity property for any $d > 0$ bond paths. In total, the complete RAC set evaluated on the five original heuristic properties consists of 30$d$+30 product-based and 12$d$ difference-based RACs for a total feature vector size of 42$d$+30. For a typical[65] cutoff of $d = 3$, this yields 156 possible features. For the mononuclear octahedral TMCs that are the focus of this work, excluding five constant features (i.e., corresponding to connectivity around the metal center) produces 151 features. In the original work, four additional descriptors of ligand denticity, oxidation state, spin state, and Hartree–Fock exchange fraction in the DFT functional were added, leading to the name RAC-155[65]. In this work, we always include the oxidation state (i.e., for 152 total features) and the spin state is included only for $\Delta E_{LD}$ (i.e., for 153 total features) but still refer to the set as RAC-155.

Because RAC heuristic properties depend on the properties of individual atoms, we revisit the extent to which such properties encode useful information about chemical similarity of distinct elements. Some, such as the Pauling electronegativity, encode close chemical similarity of distinct elements (e.g., χ: 1.88 for Co vs 1.90 for Tc). Still, the nuclear charge, $Z$, exaggerates the dissimilarity of elements with differing principal quantum number belonging to the same group (e.g., $Z$: 27 for Co vs 45 for Rh or 7 for N vs 15 for P). Thus, we expect low-complexity kernel models trained on RAC-155 to exaggerate differences in TMCs with metals or ligands from differing rows of the periodic table in comparison to those from the same row. To illustrate this, we carried out principal component analysis (PCA) on a subset (i.e., $d = 0$ or 1) of mc-RACs in RAC-155 applied to all IMP (i.e., with both 3$d$ and 4$d$ metals) complexes for which we have computed the $\Delta E_{I-L}$ property. For these TMCs, the first PC in this subset of RAC-155 primarily distinguishes the metal centers (Figure 2). Not only are 3$d$ and 4$d$ metals very distant from each other, but the strong variation of the metal $Z$ over the set in comparison to the other heuristics leads to the later 3$d$ Co being the most proximal metal to the early 4$d$ Mo.

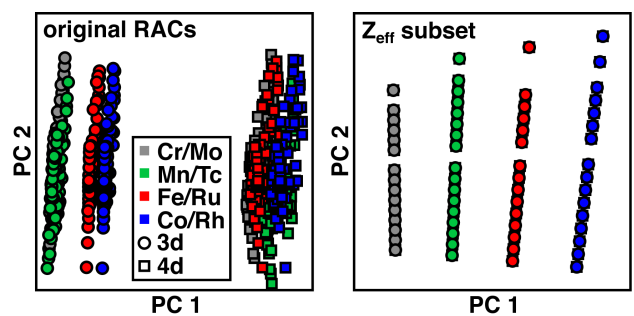

**Figure 2.** PCA obtained using the subset of features in RAC-155 that are metal-centered and $d = 0$ or 1 (left) and using only the metal-centered RACs of the $Z_{eff}$ atomic property with the same cutoff (right). The PCA was applied to 642 TMCs for which $\Delta E_{\text{I-L}}$ was computed in the IMP data set. The $3d$ TMCs circles and $4d$ TMCs squares are overlapping in the right PCA. Metals are colored according to their isovalent pairing: gray for Cr/Mo, green for Mn/Tc, red for Fe/Ru, and blue for Co/Rh, as indicated in inset legend.

We thus add to our original set of RAC heuristic properties the effective nuclear charge, $Z_{eff}$, which had been shown to be beneficial for increasing transferability in organic molecules.[80-81] This property increases the number of feature vector dimensions by $6d+6$ product-based and $3d$ difference-based RACs. For the $d = 3$ cutoff used here, this adds 33 features to our original 152- or 153-dimensional set, and we refer to this set as eRAC-185. The $Z_{eff}$ is identical for isovalent species (e.g., 9 for Co and Rh, 5 for N and P) with distinct principal quantum numbers. As expected, PCA in a subset (i.e., $d = 0$ or 1) of the new $Z_{eff}$-derived mc-RACs distinguishes TMCs by the metal $Z_{eff}$ in PC1 and by the metal-coordinating ligand atoms in PC2 (Figure 2 and Supporting Information Figure S1). As a result, isovalent $3d$ and $4d$ metals (e.g., Co and Rh) will be identical in this feature set, and the full $Z_{eff}$ set of features will order TMCs by increasing $d$ filling and groups isovalent $3d$ and $4d$ metals close to each other (Supporting Information Figures S2–S3). Conversely, both PCA and average Euclidean distances over IMP complexes for the full RAC-155 set indicate large differences between isovalent $3d$ and $4d$ metal centers and do not reproduce expected trends from the periodic table (Supporting Information Figures S2–S3). Because the new features encode intuitive relationships, we expect that incorporating them into ML property predictions should be beneficial for learning tasks that span multiple rows in the periodic table.

## 3. Methods.

### 3a. Electronic Structure Calculations.

Some of the DFT data used for training in this work was previously published in Ref. [79], and all data generation steps followed an established protocol[70] for DFT training data for octahedral, mononuclear transition-metal complexes[45], which we reiterate here. DFT geometry optimizations were performed using a development version of TeraChem v1.9.[86-87] The B3LYP[88-90] global hybrid functional was employed with the LANL2DZ[91] effective core potential for transition metals and the 6-31G* basis[92] for all other atoms. Singlet calculations were carried out in a spin-restricted formalism, while all other calculations were unrestricted. Level shifting[93] was employed to aid self-consistent field convergence with the majority-spin and minority-spin virtual orbitals each shifted by 0.25 Ha. Geometry optimizations were carried out in translation rotation internal coordinates[94] using the L-BFGS algorithm. Default tolerances of $4.5 \times 10^{-4}$ hartree/bohr and $10^{-6}$ hartree were applied in the convergence criteria for the gradient and energy difference between steps, respectively.

Initial structures were generated with molSimplify[71, 95-96], which uses OpenBabel[97-98] as a backend, and calculations were automated with molSimplify Automatic Design (mAD).[40, 70] The mAD calculations run for 24 hours of wall time with up to five resubmissions. Geometric criteria[70] were applied after each resubmission and at tighter thresholds on the final structure (Supporting Information Table S5). We use the same geometric criteria values as in ref. [70], which preserve connectivity and penalize excessive asymmetry, except we have tightened the criterion on metal–ligand distance asymmetry for homoleptic complexes (Supporting Information Table

S5). Out of 9,440 initiated 3*d* and 4*d* geometry optimizations, converged structures were obtained for 8,729 TMCs, and 6,480 TMCs satisfied all geometric criteria (Supporting Information Table S6).

As in previous work,[45, 70-71] criteria based on electronic structure properties were also used to filter the data set. For all open shell (i.e., non-singlet) complexes, calculations with deviations of $<S^2>$ from its expected $S(S+1)$ value by 1 $\mu_B^2$ or more were eliminated (196 runs) as were cases where the Mulliken spin on the metal was at least 1 $\mu_B$ lower than the expected total spin (292 runs, Supporting Information Table S6). After all filtering steps, 5,992 octahedral 3*d* or 4*d* TMCs were retained for spin-splitting property prediction, but because evaluation of this property requires the convergence of multiple spin states, the final data set sizes were smaller (Supporting Information Table S7). The $\Delta E_{LD}$ property was only evaluated for trans or homoleptic complexes with neutral ligands, and filtering steps by the $<S^2>$ and Mulliken metal spin criteria were also applied to these complexes (Supporting Information Table S8).

The retained TMCs were then processed through two additional steps to generate the data sets used in ML model training, a more restrictive procedure than that outlined in Ref.[79]. First, TMCs were paired to construct the IMP and ILP data sets, reducing the candidate TMC set sizes by a factor of two (Supporting Information Tables S3–S4). We then detected cases where the electronic structure was qualitatively distinct between isovalent pairs of complexes, as judged by the metal *d*-orbital occupations. The occupation of each *d* orbital was obtained from the NBO v6.0 package[38], and a heuristic cutoff of 3 e⁻ for the total difference summed over all *d* orbitals was used to exclude a pair of calculations (Supporting Information Figure S4). This procedure eliminates a small number (ca. 10–100 for spin-splitting and 250 for $\Delta E_{LD}$) of outliers (Supporting Information Tables S3–S4). The final IMP data sets contain ca. 300–600 pairs of spin-splitting (i.e., I-L, H-L, or H-I) $\Delta E$ values and over 1,000 pairs of $\Delta E_{LD}$ values, and the final ILP data sets are a little more than half the size of the IMP sets (Supporting Information Tables S3–S4). The total energies and structures of all TMCs converged in this work along with details of structures eliminated are provided in a .zip file in the Supporting Information.

**3b. Machine Learning.**

We largely follow a previously developed protocol[65, 70] for training kernel ridge regression (KRR) models to predict properties in open-shell transition-metal chemistry with some modifications noted. KRR models with a Gaussian kernel were trained on both RAC-155 and eRAC-185 representations and IMP or ILP data using the scikit-learn[99] software package. The pairs of complexes in the IMP or ILP sets were randomly partitioned into 80% train and 20% validation. The pairs are preserved during the random partitioning, e.g., if an IMP 3$d$ complex is in the training set, its isovalent 4$d$ counterpart will be as well. All inputs and outputs were normalized to have zero mean and unit variance on the training data.

Two types of models were trained: single-row models that only are trained on a subset of the training data (in practice, from 1–50) from the same row as the learning task, and transfer models that are supplemented with all of the training data from the alternate row. Using the 324-pair of complex IMP set for the 3$d$ to 4$d$ learning task as an example, models (see Sec. 4) were trained on a subset of the 259-complex 80% 4$d$-only (i.e., single-row models) or this same subset of points along with the 80% 3$d$ data (i.e., transfer models). Exhaustive grid search from $10^{-12}$ to $10^2$ with logarithmic spacing was used to select hyperparameters (i.e., kernel width and regularization strength). For the IMP 3$d$ to 4$d$ example, hyperparameters selected from 10-fold cross-validation errors on 3$d$ data produce models with low test set mean unsigned error (MUE) on the in-distribution 3$d$ complexes but high MUE on the out-of-distribution 4$d$ test set (i.e., the isovalent pairs to the 20% 3*d* test set, Supporting Information Figure S5). Thus, we select hyperparameters for each KRR model on the 20% data from the same row as the learning task and hereafter refer to this as the validation MUE, which provides a lower bound of a true unseen test set model error (Supporting Information Figure S5).

Feature selection was employed, as motivated by observations[65, 70] that it improves KRR model generalization in open-shell transition-metal chemistry. Feature-selected subsets were obtained in a one-shot fashion using LASSO[100] or random forest (RF)[101] as in some prior work[65]. Features with an importance greater than 1% were retained from RF. For LASSO feature selection, 5 values for the regularization strength were tested ($10^{-2}$, $10^{-1}$, $10^{0}$, $10^{1}$, or $10^{2}$), and all nonzero features were retained from the LASSO model with the lowest validation MUE. The overall feature set (i.e., from LASSO, RF, or the full feature set) that produced the lowest validation MUE was selected for training the KRR model. We report the MUE as the ensemble-average MUE and a credible interval from the ensemble standard deviation. A representative example of the selected hyperparameters, feature selection method, and MUEs is provided in detail in the Supporting Information (Supporting Information Table S9). All other model hyperparameters and feature sets are provided in the Supporting Information .zip file.

## 4. Results and Discussion.

### 4a. Correlations of Properties Between Rows.

Because our modified representations are motivated by the assumption that isovalent metals or ligands should exhibit similar properties and structure–property relationships, we first validate this expectation over the IMP and ILP sets. For the isovalent 3*d* and 4*d* transition metals in the IMP set, we observe good correlations between the adiabatic gas-phase spin-splitting energies of the first- and second-row metals (Figure 3). This correlation holds somewhat better ($R^2 = 0.9$) for the 2-electron $\Delta E_{I-L}$ and 4-electron $\Delta E_{H-L}$ than for the 2-electron $\Delta E_{H-I}$ ($R^2 = 0.65$) values (Figure 3 and Supporting Information Table S10). As observed in prior work[79], all 4$d$ TMC DFT spin-splitting energies are consistently low-spin shifted with respect to equivalent 3$d$ TMCs, especially for strong-field ligands (i.e., with positive 3$d$ TMC $\Delta E$ values). For the $\Delta E_{I-L}$ and $\Delta E_{H-L}$ cases where the correlations hold best, the 4$d$ TMC $\Delta E$ increases by around 1.4 kcal/mol for every 1 kcal/mol increase in the 3$d$ TMC (Figure 3 and Supporting Information Table S10).

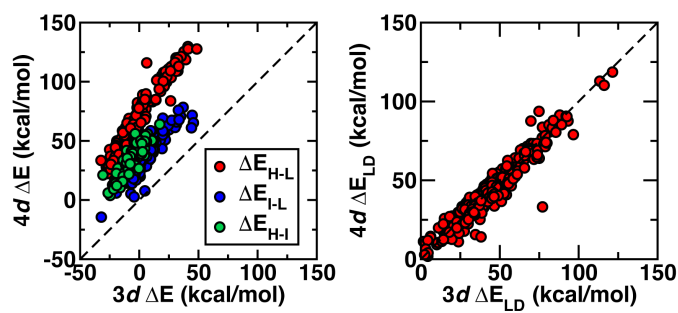

**Figure 3.** Parity plot for spin-splitting energies, $\Delta E$ (left, in kcal/mol) and ligand dissociation energies, $\Delta E_{LD}$ (right, in kcal/mol) of Fe/Ru TMCs in the IMP data set. The spin-splitting energy is colored with $\Delta E_{H-L}$ in red, $\Delta E_{I-L}$ in blue, and $\Delta E_{H-I}$ in green, and all data is shown in red at right for $\Delta E_{LD}$. The parity line shown as a dashed black line in both panes.

Despite the apparent good linear correlation over the data set, the MUE of the linear model for predicting 4$d$ $\Delta E_{H-L}$ from its 3$d$ TMC value is still large at around 10 kcal/mol, and the maximum unsigned error is even larger at 45 kcal/mol (Supporting Information Table S10). While MUEs are reduced somewhat for $\Delta E_{H-I}$ and $\Delta E_{I-L}$ (to ca. 7 kcal/mol), the range of values are also smaller for these properties. Each individual 3$d$ or 4$d$ metal/oxidation state spans a wide range of $\Delta E_{H-L}$ values (ca. 100 kcal/mol), with metal-specific distributions that reflect changes in ligand field strength in our set (i.e., from weak-field halide to strong-field methylisocyanide) for both 3$d$ and 4$d$ TMCs (Supporting Information Figure S6). When linear models are fit on each individual metal and oxidation state, the MUE of each linear model is somewhat reduced (e.g., $\Delta E_{H-L}$ 4–5 kcal/mol vs 10 kcal/mol for the full set) in part due to the fact that a smaller range of $\Delta E$ values is sampled (Supporting Information Figures S7–S9 and Tables S10–S11). Subdividing the data by metal/oxidation state and building individual linear relationships as a strategy for TMC design would nevertheless be challenged by variability in data set sizes (Supporting Information Tables S11–S12). Thus, even though these spin-splitting properties are related for metal pairs in the IMP set, their relationship is not trivial and could be expected to benefit from a flexible ML regression model. A predictive ML model would have the added benefit of not requiring the calculation of one property (e.g., 3d $\Delta E_{H-L}$) to predict the other (e.g., 4d $\Delta E_{H-L}$) that would be required for even a better linear fit.

For the ligand dissociation energy, $\Delta E_{LD}$, introduced in this work, a single linear model obtains good correlation ($R^2 = 0.91$) between 3d and 4d TMCs over all metals, oxidations, and spin states (Figure 3). The ratio of 4d:3d $\Delta E_{LD}$ values (ca. 0.93) is close to unity but indicates that ligands bind 4d TMCs somewhat more weakly than equivalent 3d TMCs (Figure 3 and Supporting Information Table S13 and Figure S10). Subdividing ligand dissociation energies by metal/oxidation state or spin provides limited benefit because subset comparisons generally have similar 4d:3d ratios and correlation coefficients to the overall set (Supporting Information Tables S13–S14 and Figure S11). Whether overall or in metal/oxidation/spin-state-specific subsets, the MUE of the linear fits is somewhat lower (4–5 kcal/mol) than was observed for spin-splitting energies (Supporting Information Tables S11 and S13).

The ILP set consists of TMCs containing ligands (i.e., six of ten) that vary by isovalent (i.e., 2p vs 3p) changes to the metal-coordinating atom of the ligand. Depending on the metal and oxidation state, the correlation of 2p and 3p spin-splitting energies ranges from good ($R^2 > 0.9$) to poor ($R^2 < 0.1$), where poorer correlations correspond most typically to unchanged spin-splitting energies with the ligand substitution and with earlier (i.e., $d^3$–$d^4$) transition metals (Supporting Information Figure S12 and Table S15). The influence of the *2p* to *3p* isovalent substitution on spin-splitting energies depends on the degree of change in ligand field strength (Figure 4). A consistent trend can be observed over all $d^5$–$d^7$ metal centers in either row for the $H_2O$/$H_2S$ and $NH_3$/$PH_3$ substitutions with a good correlation and low standard deviation (Figure 4 and Supporting Information Figure S13 and Table S16).

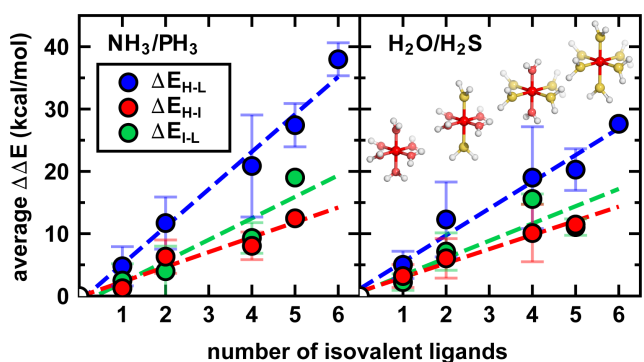

**Figure 4.** Differences in spin-splitting energies ($\Delta E_{H-L}$ in blue, $\Delta E_{H-I}$ in red, and $\Delta E_{I-L}$ in green, in kcal/mol) with isovalent ligand swaps ($NH_3$ for $PH_3$, left and $H_2O$ for $H_2S$, right) in the ILP data set averaged over all $d^5$–$d^7$ TMCs (i.e., Mn(II)/Tc(II), Fe(III/II)/Ru(III/II), and Co(III/II)/Rh(III/II)) where the spin-splitting energy is available and defined. The standard deviation is shown as an error bar along with best-fit lines also fit through (0,0).

Conversely, the limited change in field strength among the halides means these TMCs have a poorer correlation for the substitution (Supporting Information Figure S13 and Table S16). For $\Delta E_{LD}$, trends are less clear, but the *2p* to *3p* substitution of two axial ligands (i.e., including the ligand undergoing dissociation) generally decreases the ligand dissociation energy (by 12–16 kcal/mol) for the $H_2O$/$H_2S$ and $NH_3$/$PH_3$ pairs (Supporting Information Tables S17–S18). As with the IMP set, there is a clear structure to the ILP data indicating that properties of *2p* complexes are related to their *3p* counterparts, but the relationships depend more strongly on the group identity of the metal-coordinating atom.

We briefly summarize the structure–property relationships uncovered for our learning tasks to test ML model and representation generalization across the periodic table. For IMP spin-splitting energies, 3$d$/4$d$ relationships are sensitive to the spin states and ligand field strengths being compared, whereas IMP $\Delta E_{LD}$ values are comparable across 3$d$ and 4$d$ TMCs. For ILP spin-splitting energies, 3$d$/4$d$ relationships are dependent on ligand field strength, and ligand dissociation energies are more clearly influenced by the principal quantum number of the dissociated axial ligands. Given the evidence for some relationships between rows of the periodic table, ML model training is expected to benefit from the $Z_{eff}$ heuristics present in eRAC-185.

**4b. ML Property Prediction Across the Periodic Table.**

Based on the ability of linear models to predict IMP properties as well as some evidence of structure in the ILP data set, we expect that data from one row of the periodic table can be used to inform predictions on isovalent counterparts. To test this hypothesis, we construct a series of KRR models using both RAC-155 and the eRAC-185 representation. We adopt two strategies to test model generalization (see Sec. 3b). While we train all models on a randomly selected and small subset (i.e., from 10–50) of training points from the same row as the 20% validation data, we also train "transfer" models that are supplemented with all available (i.e., the full 80%) training data from the alternate row. To unambiguously identify the training data included in each model, we adopt the notation *data-representation*, where the training *data* is only from the same row (S) as the test (here, also validation) data or transfer (T) data from the alternate row is also included and *representation* is the feature set (i.e., RACs or eRACs) used by the model. For example, a T-eRAC model applied to the 3$d$ to 4$d$ $\Delta E_{H-L}$ learning task will be provided with all available 3$d$ $\Delta E_{H-L}$ data in the IMP set along with a small number (e.g., 10-50) of randomly-selected 4$d$ $\Delta E_{H-L}$ points from the IMP set and trained on eRAC-185 and feature-selected subsets (see Sec. 3b).

We first focus on 3$d$ to 4$d$ prediction of $\Delta E_{H\text{-}L}$ and $\Delta E_{LD}$ in the IMP data set to assess the relative performance of the possible data-representation combinations. Models incorporating data from both rows (i.e., T-(e)RACs) consistently outperform single-row models (i.e., S-(e)RACs), and, simultaneously, eRAC-185-trained models have lower MUEs than models trained with RAC-155 (Figure 5). The MUE is evaluated as the average error from an ensemble of 25 feature-selected models (see Sec. 3b), and observed model performance trends are typically larger in magnitude than the standard deviation of the errors from the ensemble. The benefit of eRAC-185 is most apparent when limited training data (i.e. < 20 points) from the same row as the validation data are used. For example, prediction of $\Delta E_{LD}$ from the IMP data set has a substantially lower MUE (by ca. 6.6 kcal/mol) with the T-eRAC model with 20 4$d$ points in the training set than with the S-eRAC model. When we increase the number of 4$d$ training points modestly (i.e., to 50), the relative benefit of this additional data from another row is smaller (i.e., the MUE is lower by only ca. 3.5 kcal/mol) but still significant (Supporting Information Table S19).

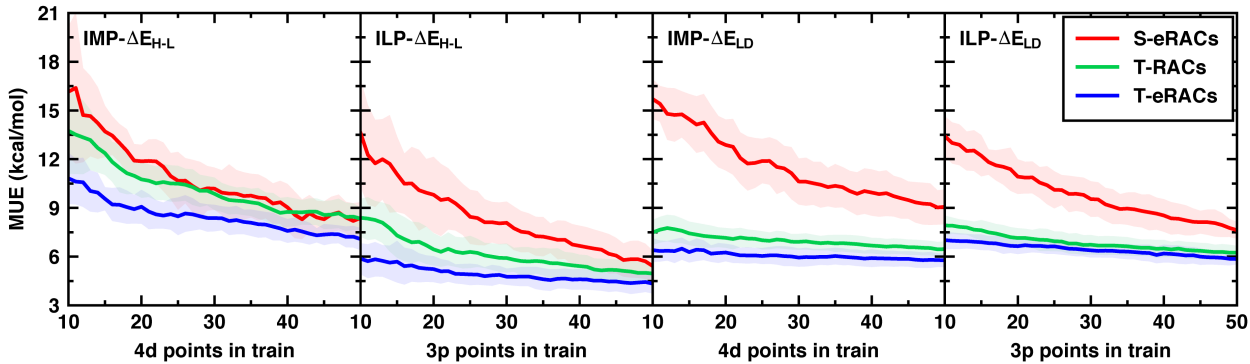

**Figure 5.** Mean unsigned error (MUE, in kcal/mol) for the prediction of $\Delta E_{H\text{-}L}$ (left) and $\Delta E_{LD}$ (right) properties for the IMP 3$d$ to 4$d$ learning task and ILP 2$p$ to 3$p$ learning task with addition of training data as indicated on the x-axis. Transfer-learning models (i.e., that contain all available 3$d$ or 2$p$ data) with eRAC-185 (blue) and RAC-155 (green) are shown alongside a single-row model (i.e., that only contain 4$d$ or 3$p$ data) trained on eRAC-185 (red). Shaded areas indicate the standard deviation of the ensemble of 25 models, and the solid line represents the average of this ensemble.

To investigate if this observed maximum benefit of eRAC-185 when few examples from the same row as the validation data are available is a general phenomenon, we expanded our analysis to models trained to predict $\Delta E_{I-L}$ and $\Delta E_{H-I}$ with only 20 4$d$ data points. Indeed, we find that transfer models consistently outperform single-row models, and the use of eRAC-185 feature set is synergistic with this benefit (Figure 6). The T-eRAC models consistently achieve the lowest overall MUEs for predicting all three spin-splitting properties, and those trained on eRAC-185 achieve larger reductions in MUE (ca. 2–5 kcal/mol) than equivalent models training using RAC-155 (ca. 1–2 kcal/mol). For predicting $\Delta E_{LD}$, transfer models significantly improve over single-row models (by ca. 6 kcal/mol), but this improvement is observed for both RAC-155 and eRAC-185 representations (Supporting Information Table S19). We attribute the difference in $\Delta E_{LD}$ model errors to the observed lack of dependence on the principal quantum number of the metal for this property (see Sec. 4a).

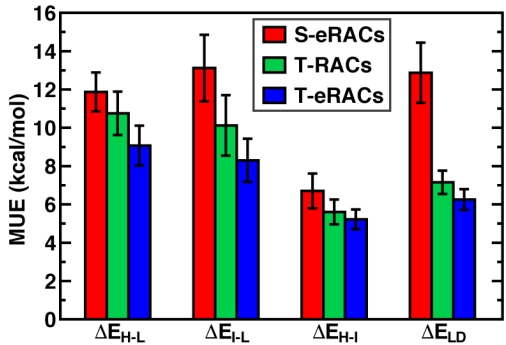

**Figure 6.** Mean unsigned error (MUE, in kcal/mol) for the prediction of 4$d$ TMC properties with only 20 4$d$ data points with eRAC-185 (red) or by including 3$d$ data with 20 4$d$ data points and using either the RAC-155 (green) or eRAC-185 (blue) on the IMP data set. The colored bars represent the average from an ensemble of 25 feature-selected KRR models, and the error bars are the standard deviation of the ensemble.

To probe the source of benefit of the inclusion of the first 20 inter-row data points, we developed an additional ML model training procedure in which we grouped inter-row data by metal in order to preferentially add one metal at a time to the training set. When the training data includes only a single 4$d$ metal (i.e. Mo, Tc, Ru, or Rh) for $\Delta E_{I-L}$, no improvement in the model is observed (Supporting Information Figure S14). Rather than reducing MUEs with more data, the MUEs are instead high and constant at approximately 15 kcal/mol when one to 15 training points of this type are added (Supporting Information Table S20). If data points are added from multiple additional 4$d$ metals, performance instead improves significantly. In this case, MUEs for predicting $\Delta E_{\text{I-L}}$ reduce to ca. 11 kcal/mol when training data includes two 4$d$ metals (Supporting Information Table S20). After adding fewer than 20 points, the T-eRAC model with knowledge of two 4$d$ metals alongside all 3$d$ metals outperforms the S-eRAC model (MUE: 13 kcal/mol) that was trained on 20 random points that included all 4$d$ metals (Figure 6 and Supporting Information Figure S15). The MUE is reduced further when a third 4$d$ metal (i.e., to 7.8 kcal/mol) and all 4$d$ metals (i.e., to 6.2 kcal/mol) are included in the training set (Supporting Information Table S20). Thus, we expect that inter-row models perform best when examples of each new metal type are included in the training data by giving the model balanced information about the relationship among all groups of the periodic table studied in the validation set. The significance of the identified sample size of 20 inter-row data points maximizing benefit in transfer models is likely attributable to reaching a sample that includes sufficient diversity of isovalent metals or ligands. This result is non-trivial because at such small set sizes, the model has information about all metals but not in combination with ligands and thus must infer the cooperativity of metal and ligand field effects.

Returning to the eRAC-185 representation, we tested whether its observed benefit in the IMP data set 3$d$ to 4$d$ learning task is general to both the reverse learning task (i.e., 4$d$ to 3$d$ IMP) as well as to changes in ligand chemistry (i.e., 2$p$ to 3$p$ ILP and 3$p$ to 2$p$ ILP). Across all of these learning tasks, a combination of transfer models and eRAC-185 reduces validation MUEs for most properties, with results and improvements comparable to those observed for the IMP 3$d$ to 4*d* learning task (Supporting Information Figure S15). In all learning tasks and properties compared, ILP 2*p* to 3*p* and 3*p* to 2*p* $\Delta E_{\text{H-I}}$ are the only cases where RAC-155-trained models outperform eRAC-185, but these difference are small (0.3 kcal/mol) and within the standard deviation (0.6 kcal/mol) of the 25-model ensembles (Supporting Information Table S21). As in the IMP data set, the benefit of inter-row, transfer models trained to predict $\Delta E_{\text{LD}}$ in the ILP set (ca. 6 kcal/mol) is observed to be independent of the representation (i.e., RAC-155 or eRAC-185) chosen. For spin-splitting energy prediction in the ILP set, we again observe that eRAC-185 improves inter-row models, with T-eRAC models outperforming single-row models by a slightly larger margin (ca. 2–6 kcal/mol) than T-RAC models (ca. 2–4 kcal/mol). Taken together, we conclude that the use of eRAC-185 in combination with a transfer model trained on a small set of examples from the new learning task provides a near-universal benefit and never significantly degrades the accuracy in comparison to a standard ML model.

**4c. Feature Analysis Reveals Learned Chemical Trends.**

To investigate the source of the improved performance observed when using eRAC-185, we evaluate the averaged, down-selected feature space. For each model in the ensemble, features are selected independently (i.e., from one-shot LASSO or random forest 1% cutoff, see Sec. 3b), so this analysis is on the normalized feature contribution is averaged over the KRR models in the ensemble. Features containing the $Z_{\text{eff}}$ atomic property are frequently selected in inter-row models for predicting both spin-splitting energies and $\Delta E_{\text{LD}}$ (see Supporting Information .zip file for all features and models). For 3*d* to 4*d* prediction of $\Delta E_{\text{I-L}}$ (i.e., on the IMP set), $Z_{\text{eff}}$ eRACs involving the metal center, atoms one bond-path away from the metal, and atoms two bond-paths away from the metal are all selected (Figure 7). This suggests that the T-eRAC models achieve performance improvements by positioning both metals and ligands closer to their inter-row counterparts in the feature space than standard RACs. In contrast, for 3$d$ to 4$d$ prediction of $\Delta E_{LD}$, $Z_{eff}$ features involving the metal are rarely selected (Figure 7). This analysis mirrors our observation that the expanded representation did not lead to significant improvements for a $\Delta E_{LD}$ inter-row, transfer model (see Sec. 4b) due to the relative lack of dependence of the ligand dissociation on the principal quantum number of the metal.

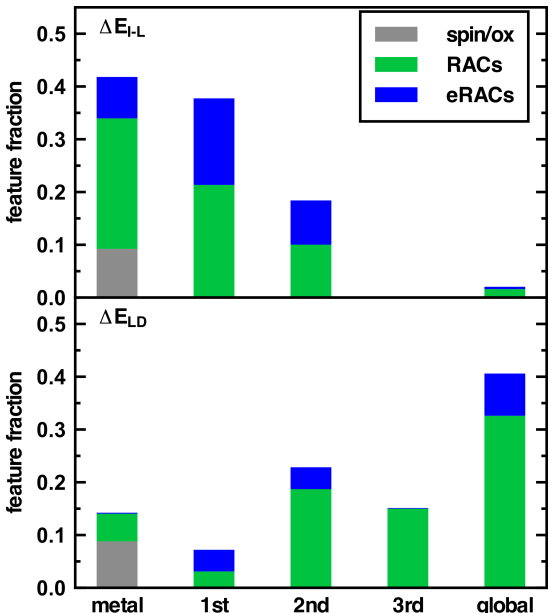

**Figure 7.** Feature fraction grouped by the most metal-distant atom in the feature (i.e., metal, 1$^{st}$, 2$^{nd}$, 3$^{rd}$, or global) and shown as a stacked bar plot. Colors within each feature fraction correspond to whether the features are obtained from RACs (green), the eRAC-specific $Z_{eff}$ features (blue), or features encoding the metal center's spin and oxidation state (gray). The selected features are obtained from their normalized weight in an ensemble of 25 models (i.e., T-RACs or T-eRACs) trained to predict 4$d$ properties using 3$d$ data along with 20 4$d$ data points.

We analyzed other learning tasks (i.e., beyond 3$d$ to 4$d$ learning of $\Delta E_{I-L}$ and $\Delta E_{LD}$) to confirm that the benefit of using eRAC-185 was due to the presence of $Z_{eff}$ features in feature-selected subsets. Prior to feature selection, the $Z_{eff}$ features comprise 18% of the initial eRAC-185 set, and they are frequently retained after feature selection regardless of the learning task or property (Supporting Information Figure S16). For the $\Delta E_{I-L}$ and $\Delta E_{H-I}$ properties where $Z_{eff}$-based features were most beneficial in model performance (see Sec. 4b), $Z_{eff}$ contributions are enriched (to ca. 23% of features) after feature selection. For $\Delta E_{LD}$ and $\Delta E_{H-I}$, cases where $Z_{eff}$-based features provided less benefit, they are retained at a comparable frequency (ca. 18% of the down-selected features) to others in the set, confirming that these features are neither particularly detrimental nor beneficial (Supporting Information Table S22).

As in previous work[38, 65, 70, 72] we also analyzed the selected features to reveal qualitative chemical trends in the data set. We select $\Delta E_{I-L}$ $3d$ to $4d$ learning as a representative property for spin-splitting energy trends (Supporting Information Figure S16). Features retained by models trained on $\Delta E_{I-L}$ are strongly metal-local, with 80% corresponding to the metal or its immediate coordinating atoms. Since this result recapitulates observations from prior work[65] that emphasized the metal-local nature of spin-splitting energy prediction, the tailored representation preserves known trends (Figure 7 and Supporting Information Table S23). We next analyze the features selected for predicting $\Delta E_{LD}$, which is a property for which we had not previously trained ML models. In this case, the retained features are 79% distal (i.e., with information from atoms two or more bonds away) from the metal (Figure 7 and Supporting Information Table S23). The high weight of ligand-based features and more global information is consistent with our observations of reduced dependence on the principal quantum number of the metal for $\Delta E_{LD}$ prediction (Figure 3). Thus, feature analysis suggests it is possible to design ligand binding strength (e.g., for catalysis) in a metal-row-independent fashion (e.g., predicting $4d$ properties from data obtained on $3d$ TMCs). Conversely, because spin state is dictated both by the metal center and only the immediate coordinating atoms of the ligand, inter-row transfer models trained on eRACs and minimal but chemically diverse TMCs[74] should enable accelerated prediction of $4d$ TMC spin-splitting in cases where only $3d$ TMC data is available.

Because the property prediction obtained from a KRR model depends on a combination of feature-space distances and the kernel width, the relative distances between TMCs in a representation are critical for understanding model performance. We thus quantify the Euclidean norm distance between complexes with differing metals in feature-selected subspaces that have been averaged and normalized over the 25-model ensembles used for property prediction. In the features selected in T-RAC models, these Euclidean distances do not represent what we would expect from the arrangement of elements on the periodic table (Supporting Information Table S24). For example, in the 3$d$ to 4$d$ prediction of $\Delta E_{\text{I-L}}$, Cr is positioned closer to Fe and Co in the feature space than it is to the more chemically similar Mn. Furthermore, Cr is comparatively distant to all of the 4$d$ metals, artificially reducing its influence on predicting their properties (Figure 8). In contrast, the features retained in T-eRAC models rearrange the feature space to more closely match intuition from the periodic table, with Cr in closest proximity to Mn and Mn in closest proximity to Cr and Fe. The 3$d$ and 4$d$ metal pairs are also shifted closer together in T-eRAC models. For example, 3$d$ Cr is closer to isovalent 4$d$ Mo than to a dissimilar 3$d$ metal such as Co (Figure 8). The rearrangement of the feature space is observed for other metals, with 3$d$/4$d$ pairs shifted an average of 13% closer in T-eRAC models in comparison to equivalent T-RAC models (Supporting Information Table S24). These results suggest that the reduced prediction error when using eRAC-185 (see Sec. 4b) can be attributed to this representation encoding chemical trends more efficiently.

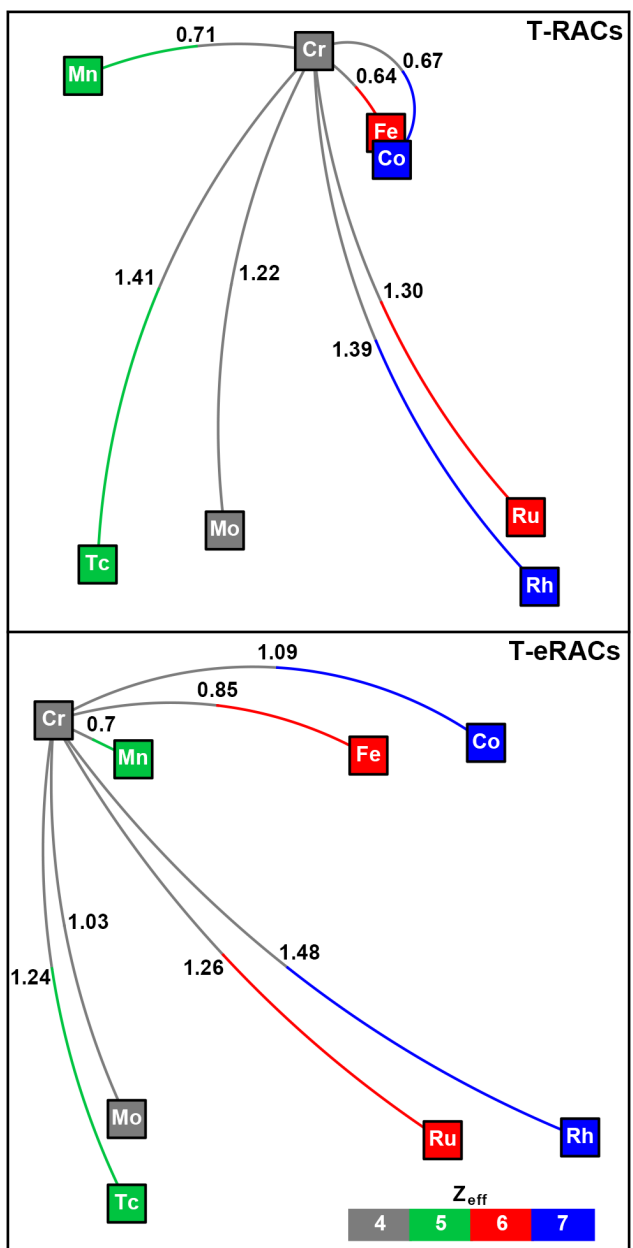

**Figure 8.** Distribution of average, normalized distances of TMCs with the feature-selected subsets of T-RACs (top) or T-eRACs (bottom) from the ensemble of 25 KRR models trained to predict 4$d$ $\Delta E_{\text{I-L}}$ based on 3$d$ data and 20 4$d$ examples (i.e., transfer models) visualized in a two-dimensional projection of the feature space. The average distances between Cr TMCs and TMCs with other metal centers are labeled. Elements are colored by their $Z_{\text{eff}}$ values to highlight isovalent 3$d$/4$d$ pairs according to inset legend. The feature-space distances were normalized such that the average distance between any two TMCs is equal to one, and the weight of features selected from the ensemble were also normalized.

## 5. Conclusions.

Since chemical relationships between neighboring elements in the periodic table can be expected, efficient strategies for ML-driven discovery should leverage representations that preserve these similarities. After confirming the strong inter-row structure–property relationships for spin-splitting energies and ligand dissociation in isovalent 3$d$/4$d$ metals and 2$p$/3$p$ ligands in large sets (ca. 1000 pairs) of data, we identified a strategy for increasing ML model transferability. We introduced eRAC-185, a tailored representation that blends together the monotonically increasing nuclear charge with the periodic in nature effective nuclear charge heuristics. We showed how the addition of the $Z_{eff}$ heuristic property in eRACs altered feature-space distances to encode a more intuitive degree of similarity among isovalent elements essential for kernel-based ML models.

Next, we demonstrated the synergistic value of eRACs alongside a transfer learning approach to leverage inter-row data in sets containing isovalent metals or ligand-coordinating atoms in transition-metal chemistry. To assess the potential improved performance of eRAC-185, we trained KRR models using data from either only the same row as the prediction task or both rows of the periodic table. While including data from the alternate row always reduced prediction errors, the eRAC-185 model errors were most substantially improved in the limit of very low data (ca. 20 points). The model error improvement was strongest when the relationship being learned deviated most from parity. This highlights how transferable representations will be essential to improve ML model generalization from one row of the periodic table to another especially when carrying out chemical discovery where limited prior knowledge is available.

To identify differences in models trained on eRAC-185, we analyzed the distance and heuristic atomic properties retained during feature selection. Feature-space distances of TMCs grouped and averaged by metals confirmed that eRAC-185 subsets faithfully represented trends expected from the periodic table, causing elements from the same or adjacent group to influence KRR predictions most strongly. These representations, models, and approach can be leveraged to accelerate the discovery of earth-abundant catalysts or materials from known materials featuring scarcer elements. We expect our strategies for inter-row design and expanding transferability to be broadly useful to other materials challenges across the periodic table.

ASSOCIATED CONTENT

**Supporting Information**.
Spin multiplicity definitions; subset of combinations studied for cis TMCs; statistics for filtering and pairing structures as well as 3d/4d comparisons to obtain properties; PCA of RAC and eRAC representations; comparison of Euclidean distance between metals by featurization type; geometry metrics used for geometric filtering; number of compounds failed at each filtering step with reasons; details of NBO d orbital check elimination; cross-validation heat maps for hyperparameter selection; KRR model hyperparameters by model number; linear models for inter-row property prediction for the IMP dataset, overall and broken down by metal; histograms of property distributions for spin splitting energy; inter-row parity plots for properties by metal and oxidation state; ligand dissociation energy distribution broken down by metal and oxidation state; frequencies of ligand dissociation energy by metal, oxidation, and spin state in the IMP and ILP data sets; parity plots for spin splitting energy in the IMP set; changes in property by ligand mixing for three spin splitting energy properties; linear models for spin splitting energies in ILP dataset split by metal, oxidation state, and type of spin splitting energy; change in spin splitting energy as a function of ligand mixing in the ILP data set; quantification of ligand substitution in ILP dataset quantified by ligand type; differences in ligand dissociation energy for compounds that vary only in equatorial ligands or only in axial ligands; quantification of test set errors for KRR model ensembles for various inter-row learning tasks; model performance by ordered metal addition for 4$d$ metals; quantification of errors with ordered metal addition of 4$d$ metals; mean unsigned errors for various inter-row learning schemes over the IMP and ILP data sets; stacked bar charts for feature character for different properties obtained during each scheme of inter-row learning, and tabulation of corresponding feature character by property and bond depth; quantification of feature space distances for feature selected subspaces of 3$d$/4$d$ models in the IMP data set (PDF)
Structures and total energies of all 3$d$ and 4$d$ TMCs studied in this work; reasons complexes were not included in the dataset; spin splitting values for all pairs of 3$d$/4$d$ TMCs; ligand dissociation energies for 3$d$/4$d$ TMCs; total energies of all TMCs; KRR model performance and selected features and ensemble quantification (ZIP)

This material is available free of charge via the Internet at http://pubs.acs.org.

AUTHOR INFORMATION

**Corresponding Author**

*email: hjkulik@mit.edu phone: 617-253-4584

**Notes**

The authors declare no competing financial interest.

ACKNOWLEDGMENT

The authors acknowledge initial partial support in data generation steps and machine learning respresentation development (for A.N.) by the United States Department of Energy under grant number DE-SC0012702. The work was also supported by the Office of Naval Research under grant numbers N00014-17-1-2956 and N00014-18-1-2434 (D.H. and J.P.J.) as well as N00014-20-1-2150 (C.D., D.H., and N.A.) as well as DARPA grant D18AP00039 (for N.A., C.D., and A.N). Ligand dissociation studies were supported by the National Science Foundation under grant number CBET-1704266 (for A.N.). A.N. was partially supported by a National Science Foundation Graduate Research Fellowship under Grant #1122374. H.J.K. holds a Career Award at the Scientific Interface from the Burroughs Wellcome Fund and an AAAS Marion Milligan Mason Award, which supported this work. This work was carried out in part using computational resources from the Extreme Science and Engineering Discovery Environment (XSEDE), which is supported by National Science Foundation grant number ACI-1548562. The authors thank Adam H. Steeves for providing a critical reading of the manuscript.

**For Table of Contents Use Only**

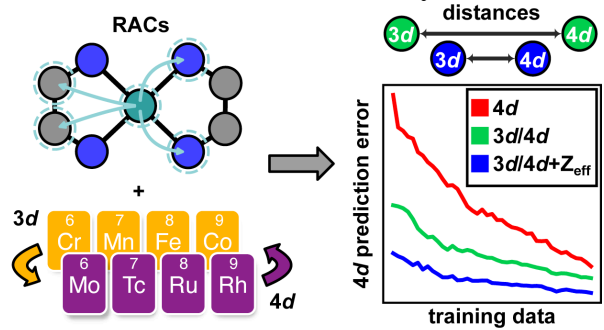

# Supporting Information for
## *Representations and Strategies for Transferable Machine Learning Models in Chemical Discovery*

Daniel R. Harper[1,2,#], Aditya Nandy[1,2,#], Naveen Arunachalam[1], Chenru Duan[1,2], Jon Paul Janet[1], and Heather J. Kulik[1,*]

[1]Department of Chemical Engineering, Massachusetts Institute of Technology, Cambridge, MA 02139

[2]Department of Chemistry, Massachusetts Institute of Technology, Cambridge, MA 02139

[#]These authors contributed equally.

**Content**

**Table S1.** Definition of spin state multiplicities (2*S*+1) studied for complexes grouped by the formal *d* electron count of the transition metal. Only LS and IS states are defined for the $d^3$ and $d^7$ complexes.

| $d^x$ | 3d M$^{(ox)}$ | 4d M$^{(ox)}$ | LS | IS | HS |
|---|---|---|---|---|---|
| $d^3$ | Cr$^{III}$ | Mo$^{III}$ | doublet | quartet | -- |
| $d^4$ | Cr$^{II}$, Mn$^{III}$ | Mo$^{II}$, Tc$^{III}$ | singlet | triplet | quintet |
| $d^5$ | Mn$^{II}$, Fe$^{III}$ | Tc$^{II}$, Ru$^{III}$ | doublet | quartet | sextet |
| $d^6$ | Fe$^{II}$, Co$^{III}$ | Ru$^{II}$, Rh$^{III}$ | singlet | triplet | quintet |
| $d^7$ | Co$^{II}$ | Rh$^{II}$ | doublet | quartet | -- |

**Table S2.** Ligand combinations enumerated for the cis arrangement (i.e., M(L1)$_4$(L2)$_2$) in the final data set. Each ligand combination is used with 12 metal and oxidation state combinations: Cr$^{II}$/Mo$^{II}$, Mn$^{II/III}$/Tc$^{II/III}$, Fe$^{II/III}$/Ru$^{II/III}$, and Co$^{III}$/Rh$^{III}$ in 3 possible spin states (i.e., no Cr$^{III}$/Mo$^{III}$ and Co$^{II}$/Rh$^{II}$ which are only in two spin states) for a total of 36 TMCs per ligand combination. This leads to 1080 TMCs with ligands in the cis arrangement considered.

| # | L1 (eq. and ax.) | L2 (eq.) |
|---|---|---|
| 1 | acetonitrile | methyl isocyanide |
| 2 | ammonia | acetonitrile |
| 3 | ammonia | carbonyl |
| 4 | ammonia | chloride |
| 5 | ammonia | cyanide |
| 6 | ammonia | fluoride |
| 7 | ammonia | hydrogen sulfide |
| 8 | ammonia | methyl isocyanide |
| 9 | ammonia | phosphine |
| 10 | ammonia | water |
| 11 | carbonyl | cyanide |
| 12 | chloride | hydrogen sulfide |
| 13 | chloride | phosphine |
| 14 | cyanide | carbonyl |
| 15 | fluoride | ammonia |
| 16 | fluoride | water |
| 17 | hydrogen sulfide | chloride |
| 18 | hydrogen sulfide | phosphine |
| 19 | methyl isocyanide | acetonitrile |
| 20 | phosphine | chloride |
| 21 | phosphine | hydrogen sulfide |
| 22 | water | acetonitrile |
| 23 | water | ammonia |
| 24 | water | carbonyl |
| 25 | water | chloride |
| 26 | water | cyanide |
| 27 | water | fluoride |
| 28 | water | hydrogen sulfide |
| 29 | water | methyl isocyanide |
| 30 | water | phosphine |

**Table S3.** Isovalent metal pairing (IMP) data set sizes for each property studied. From left to right: starting from the filtered total (i.e., retained TMCs), if a 3$d$ or 4$d$ TMC point is missing from a pair or if a pair fails the electronic structure (ES) check (i.e., from deviations in the $d$ orbital populations), the points are eliminated. The final number of points and pairs (shown in parentheses) in each set is indicated.

| Property | Filtered total (3$d$ and 4$d$ TMCs) | 3$d$ or 4$d$ missing from pair | eliminated by ES check | Final # (# pairs) |
|---|---|---|---|---|
| $\Delta E_{H-L}$ | 1,030 | 374 | 8 | 648 (324) |
| $\Delta E_{I-L}$ | 1,853 | 507 | 86 | 1,260 (630) |
| $\Delta E_{H-I}$ | 961 | 389 | 18 | 554 (277) |
| $\Delta E_{LD}$ | 2,788 | 430 | 244 | 2114 (1057) |

**Table S4.** Isovalent ligand pairing (ILP) data set sizes for each property studied. Starting from the filtered (i.e., retained TMCs), if a 3$d$ or 4$d$ TMC point is missing from a pair or if a pair fails the electronic structure (ES) check, (i.e., from deviations in the $d$ orbital populations), the points are eliminated. The final number of points and pairs (shown in parentheses) in each set is indicated.

| Property | Filtered total (3$d$ and 4$d$ TMCs) | missing from isovalent ligand pair | eliminated by E.S. check | Final # (# pairs) |
|---|---|---|---|---|
| $\Delta E_{H-L}$ | 1,030 | 630 | 40 | 360 (180) |
| $\Delta E_{I-L}$ | 1,853 | 1,053 | 158 | 642 (321) |
| $\Delta E_{H-I}$ | 961 | 609 | 16 | 336 (168) |
| $\Delta E_{LD}$ | 2,788 | 1,452 | 244 | 1,092 (546) |

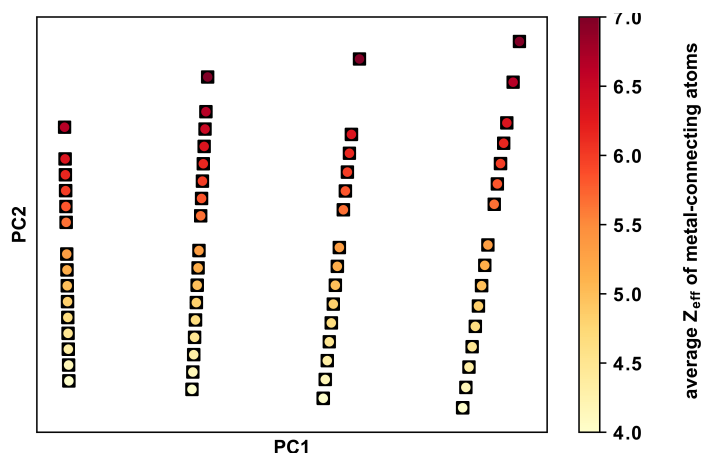

**Figure S1.** Principal component analysis of metal-centered features constructed from a subset of $Z_{eff}$ mc-RACs (i.e., $d$ = 0 or 1). The 1853 data points (i.e., from IMP $\Delta E_{I-L}$) are colored based on the average $Z_{eff}$ of metal-connecting ligand atoms (many points overlap). The first PC differentiates TMCs based on the group number of the metal atom, and the second PC by the group number of the metal-connecting ligand atom.

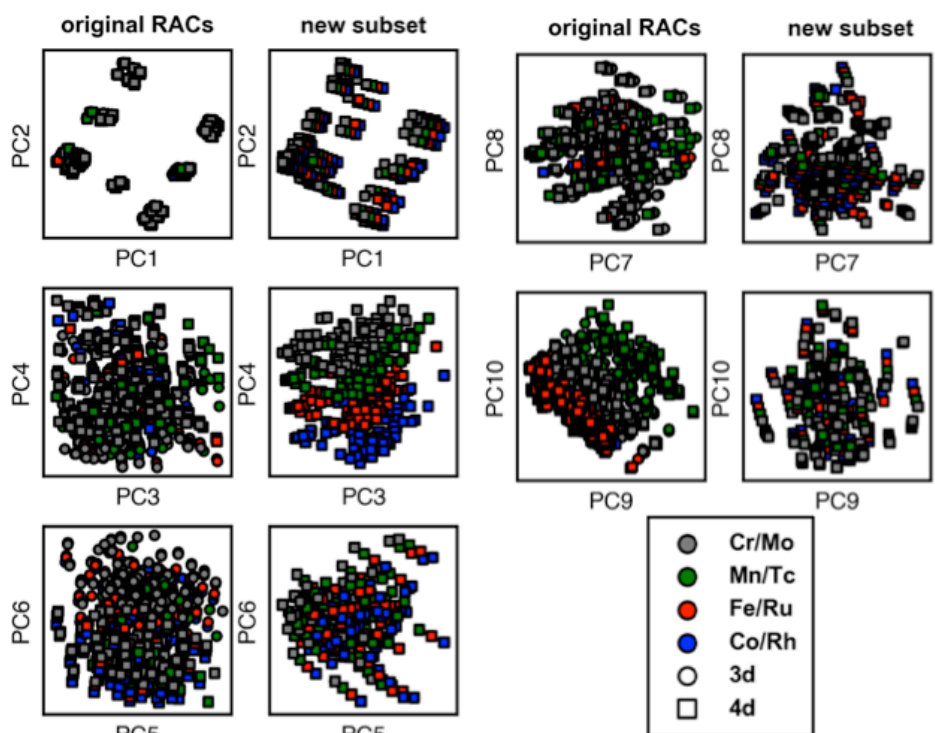

**Figure S2.** PCA of the $\Delta E_{I-L}$ points in the IMP data set for the RAC-155 representation (original RACs, columns 1 and 3) and the 33 RACs based on the $Z_{eff}$ atomic property (new subset, columns 2 and 4). All points are colored according to inset legend. The symbol indicates the metal center's row on the periodic table, and *3d* and *4d* points overlap in the new subset representation. The fourth PC highlights the key distinction between original RACs and the new features.

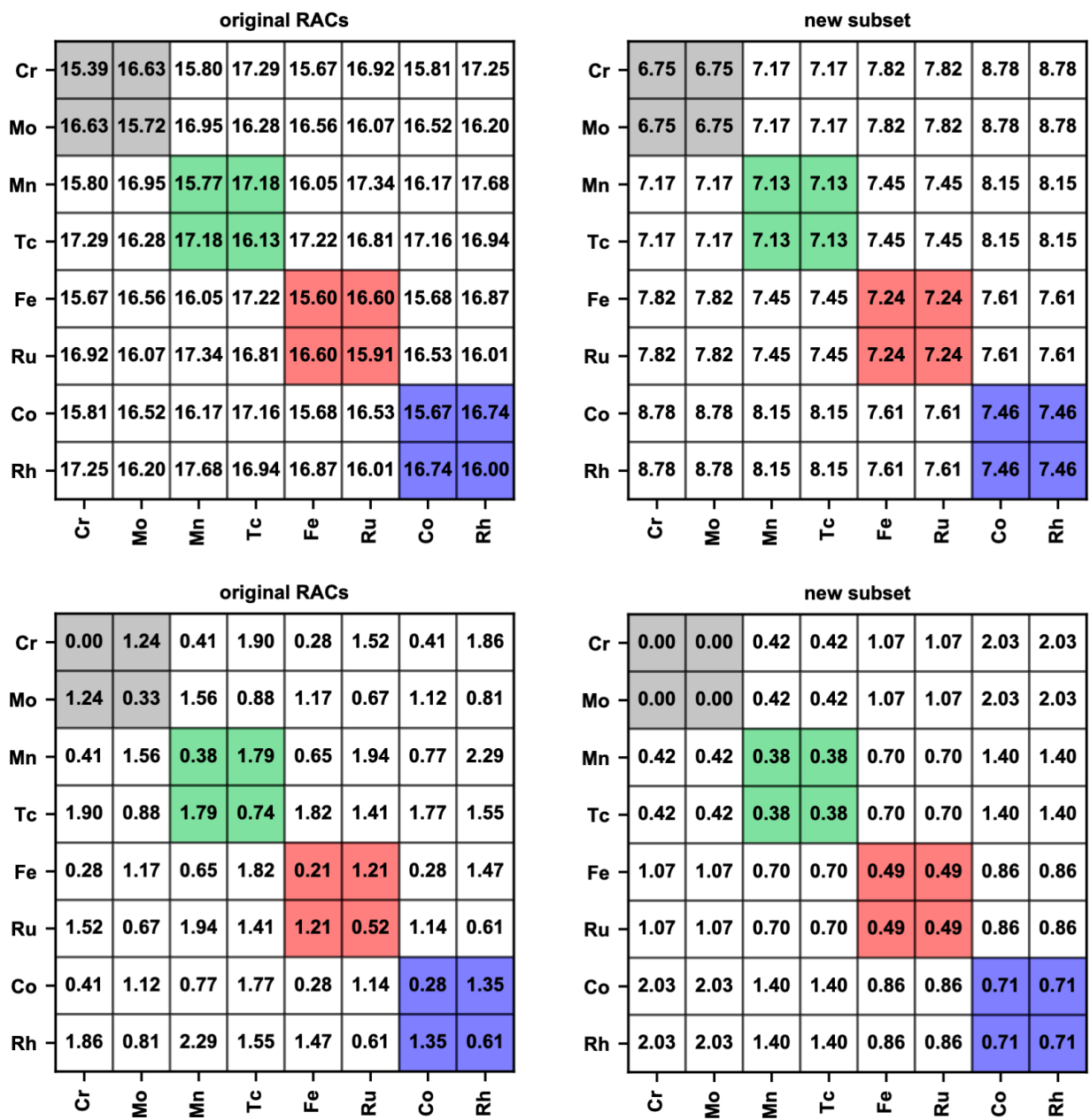

**Figure S3.** Average Euclidean distance (top) among complexes in the $\Delta E_{\text{I-L}}$ portion of the IMP data set grouped by metal along with the average Euclidean distance after subtracting the minimum value in the matrices (bottom). These properties are shown for the RAC-155 representation (left) along with the 33 added RACs with $Z_{\text{eff}}$ (right).

**Table S5.** Geometry check parameters employed to determine if calculations converged to a suitable octahedral geometry. Cutoffs for final calculations, along with loose cutoffs employed during resubmission (indicated in parentheses) are listed: coordination number of 6 (i.e., as judged through interatomic distances being within 1.37 Å x the sum of the respective elements' covalent radii); first coordination sphere metrics including the mean and maximum (max.) deviation in the angle ($\Delta\theta(C_i\text{-}M\text{-}C_j)$) formed with coordinating atoms ($C_i$ or $C_j$) and the metal from the expected values of 90° or 180° as well as the maximum overall difference between metal-coordinating atom bond lengths over all ligands in the equatorial (eq.) plane; ligand distortion metrics including the maximum root mean square deviation (RMSD) of any atom from the starting structure. For ligands that are expected to be linear, additional checks are applied on the deviation of the angle formed by the metal and the first two atoms of the ligand (A, B) from 180°. Tightened checks for homoleptic TMCs are indicated for max($\Delta d$) with a special threshold for singlet states due to our expectation of higher symmetry for these structures.

| Coordination number | | | |
|---|---|---|---|
| 6 (6) | | | |
| **First coordination sphere** | | | |
| mean($\Delta\theta(C_i\text{-}M\text{-}C_j)$) 12° (16°) | max($\Delta\theta(C_i\text{-}M\text{-}C_j)$) 22.5° (27.0°) | max($\Delta d$) 1.00 Å (1.25 Å) homoleptic: 0.4 Å (1.25 Å) homoleptic singlets: 0.2 Å (1.25 Å) | max($\Delta d_{eq}$) 0.35 Å (0.45 Å) |
| **Ligand distortion metrics** | | | |
| max(RMSD) 0.30 Å (0.40 Å) | | mean($\Delta\theta(M\text{-}A\text{-}B)$) 20° (30°) | max($\Delta\theta(M\text{-}A\text{-}B)$) 28° (40°) |

**Table S6.** Number of data points for $3d$ or $4d$ TMCs eliminated at each sequential filtering step. For both $3d$ and $4d$ TMCs, the initial pool of candidates, number of points retained after each filtering step, and the number of points eliminated at that step are all listed. Cumulative totals of eliminated complexes are listed as "overall." The filtering steps were: i) checking for convergence (completion, passing loose thresholds upon five 24-hour job resubmissions), ii) checking geometries against defined metrics, with different criteria for heteroleptic and homoleptic complexes, iii) checking for deviations of $<S^2>$ from the expected value by more than 1 $\mu_B^2$, and iv) checking for deviations of metal spin from the total spin of more than 1 $\mu_B$.

| Check Performed | 3d TMCs | | 4d TMCs | |
|---|---|---|---|---|
| | Retained | Eliminated at step | Retained | Eliminated at step |
| Start | 4,720 | -- | 4,720 | -- |
| Converged | 4,282 | 418 | 4,447 | 273 |
| Geometry (Heteroleptic) | 3,487 | 795 | 3,028 | 1,419 |
| Geometry (Homoleptic) | 3,479 | 8 | 3,001 | 27 |
| $<S^2>$ | 3,335 | 144 | 2,939 | 62 |
| Metal spin | 3,197 | 138 | 2,795 | 144 |
| **Overall** | **3,197** | **1,523** | **2,795** | **1,925** |

**Table S7.** From the converged geometry optimizations that pass all checks, the number of pairs of $\Delta E_{\text{H-L}}$, $\Delta E_{\text{I-L}}$, and $\Delta E_{\text{H-I}}$ adiabatic spin-splitting energies that can be evaluated for $3d$ and $4d$ TMCs (i.e., both spin states converged and passed filtering checks). For the earliest Cr(III)/Mo(III) and latest Co(II)/Rh(II) metal/oxidation state pairs, only $\Delta E_{\text{I-L}}$ is defined.

| Category | $\Delta E_{\text{H-L}}$ | $\Delta E_{\text{I-L}}$ | $\Delta E_{\text{H-I}}$ |
|---|---|---|---|
| All converged points | 3163 | 4408 | 2652 |
| -Missing one point | 1103 | 702 | 730 |
| Final $\Delta E$ pairs | 1030 | 1853 | 961 |

**Table S8.** Number of $\Delta E_{\text{LD}}$ $3d$ or $4d$ TMC data points eliminated after filtering on deviations of $<S^2>$ or the metal center spin of the TMC with the ligand dissociated (dissoc. TMC). We evaluate this property for the subset of TMCs with trans and homoleptic symmetry to avoid uncertainty about which axial ligand to dissociate and only study cases with neutral ligands to avoid changing charge of the TMC upon ligand removal.

| Filtering step | $3d$ TMCs | | $4d$ TMCs | |
|---|---|---|---|---|
| | Retained | Eliminated at step | Retained | Eliminated at step |
| Retained octahedral TMCs | 3,197 | -- | 2,795 | -- |
| Dissoc. TMC $<S^2>$ | 3,156 | 41 | 2,767 | 28 |
| Dissoc. TMC M spin | 3,140 | 16 | 2,398 | 369 |
| Trans and homoleptic TMCs only | 1,934 | 1,206 | 1,659 | 739 |
| Neutral ligands only | 1,500 | 434 | 1,288 | 371 |
| **Overall** | **1,500** | **1,697** | **1,288** | **1,507** |

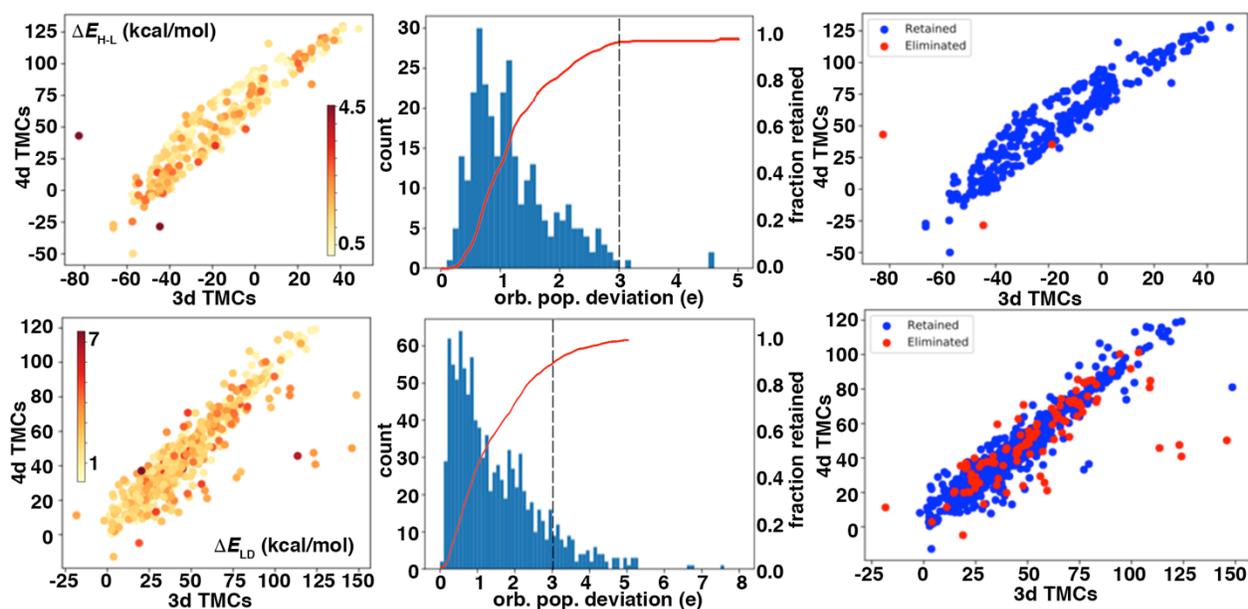

**Figure S4.** The heuristic cutoff for the orbital population (orb. pop.) deviation check was chosen based on the relationships between differences and properties for $\Delta E_{H-L}$ and $\Delta E_{LD}$ in the IMP data set. The difference in *d*-orbital occupations from NBO analysis is shown (left) and colored according to the inset color bars. The chosen value (3 e$^-$) eliminates cases where the deviation in orbital populations is most extreme (middle pane, fraction retained shown as red line with right axis). The right pane shows the data distinguishing the points that are eliminated (red) or retained (blue) using our heuristic cutoff 3 e$^-$, with both properties shown in kcal/mol.

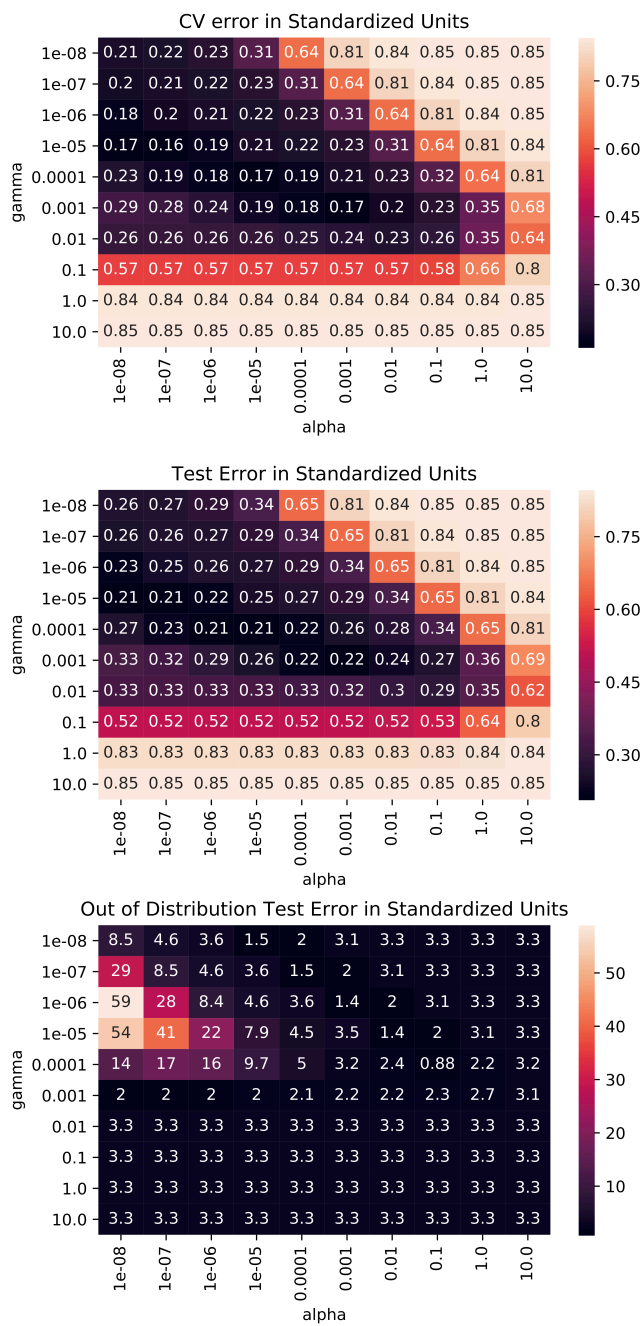

**Figure S5.** Evaluation of KRR model errors for the 324-complex-pair IMP dataset for $\Delta E_{\text{H-L}}$ with eRAC-185. The model was trained only on 80% $3d$ data, and an in-distribution test set is the remaining 20% of the $3d$ data. An out-of-distribution test set is the 20% $4d$ data that correspond to the isovalent pairs to the in-distribution 20% $3d$ test data. The comparison of error evaluations includes 10-fold cross validation (top) on the $3d$ complexes in training, error on the in-distribution test set (middle), and an out of distribution test set (bottom). In all cases, the comparisons are for α and γ hyperparameters from a grid search. Since CV errors do not reflect the out-of-distribution test set errors, we select hyperparameters directly based on the validation set in this work.

**Table S9.** Hyperparameters (α and γ) selected for an ensemble of 25 KRR models predicting the $\Delta E_{\text{I-L}}$ property (524 3$d$ TMC points in train) in the IMP set when 20 4$d$ TMC points are included in the training set. All models use the eRAC-185 representation as a starting point and retain the features in this case from a 1% cutoff of a random forest model (RF_cutoff). The test MUE (on the 20% set aside 4$d$ TMC points) of each model is reported along with ensemble average and standard deviation over all 25 models. Hyperparameters for models predicting other properties, using different data sets, and including a different number of training points from the same row as the validation data are provided in the Supporting Information .zip as .csv files.

| Model Number | Feature Selection | α | γ | Test MUE (kcal/mol) |
|---|---|---|---|---|
| 1 | RF_cutoff | 1.00E-11 | 1.00E-07 | 7.59 |
| 2 | RF_cutoff | 1.00E-11 | 1.00E-06 | 7.09 |
| 3 | RF_cutoff | 1.00E-05 | 1.00E-04 | 7.14 |
| 4 | RF_cutoff | 1.00E-04 | 1.00E-03 | 7.07 |
| 5 | RF_cutoff | 1.00E-08 | 1.00E-04 | 8.27 |
| 6 | RF_cutoff | 1.00E-07 | 1.00E-04 | 7.27 |
| 7 | RF_cutoff | 1.00E-03 | 1.00E-02 | 8.76 |
| 8 | RF_cutoff | 1.00E-11 | 1.00E-07 | 9.72 |
| 9 | RF_cutoff | 1.00E-05 | 1.00E-03 | 9.24 |
| 10 | RF_cutoff | 1.00E-03 | 1.00E-03 | 9.07 |
| 11 | RF_cutoff | 1.00E-04 | 1.00E-03 | 8.09 |
| 12 | RF_cutoff | 1.00E-02 | 1.00E-02 | 6.93 |
| 13 | RF_cutoff | 1.00E-07 | 1.00E-04 | 8.82 |
| 14 | RF_cutoff | 1.00E-11 | 1.00E-06 | 8.09 |
| 15 | RF_cutoff | 1.00E-05 | 1.00E-04 | 7.71 |
| 16 | RF_cutoff | 1.00E-09 | 1.00E-05 | 7.52 |
| 17 | RF_cutoff | 1.00E-03 | 1.00E-03 | 10.91 |
| 18 | RF_cutoff | 1.00E-08 | 1.00E-04 | 9.41 |
| 19 | RF_cutoff | 1.00E-05 | 1.00E-04 | 8.53 |
| 20 | RF_cutoff | 1.00E-03 | 1.00E-02 | 8.22 |
| 21 | RF_cutoff | 1.00E-05 | 1.00E-03 | 7.43 |
| 22 | RF_cutoff | 1.00E-11 | 1.00E-07 | 8.55 |
| 23 | RF_cutoff | 1.00E-11 | 1.00E-06 | 10.97 |
| 24 | RF_cutoff | 1.00E-12 | 1.00E-07 | 7.95 |
| 25 | RF_cutoff | 1.00E-06 | 1.00E-04 | 9.06 |
| Ensemble | | | | 8.38±1.11 |

**Table S10.** Linear models (y = ax+b, a, unitless, and b, in kcal/mol) for properties of 3$d$ TMCs (x) to 4$d$ TMCs (y) in the IMP data set (set sizes shown as # points). The mean unsigned error (MUE) in predicting the 4$d$ metal property from the 3$d$ metal property using the linear model is reported along with the number of points in the set. Lines were fit using ordinary least squares, and the $R^2$ value for each line is shown.

| Property | a | b | $R^2$ | MUE (kcal/mol) | # points |
|---|---|---|---|---|---|
| $\Delta E_{H-L}$ | 1.41 | 75.65 | 0.89 | 9.69 | 324 |
| $\Delta E_{I-L}$ | 1.38 | 27.45 | 0.91 | 7.06 | 630 |
| $\Delta E_{H-I}$ | 0.97 | 40.96 | 0.65 | 6.14 | 277 |
| $\Delta E_{LD}$ | 0.93 | 3.75 | 0.91 | 4.88 | 1057 |

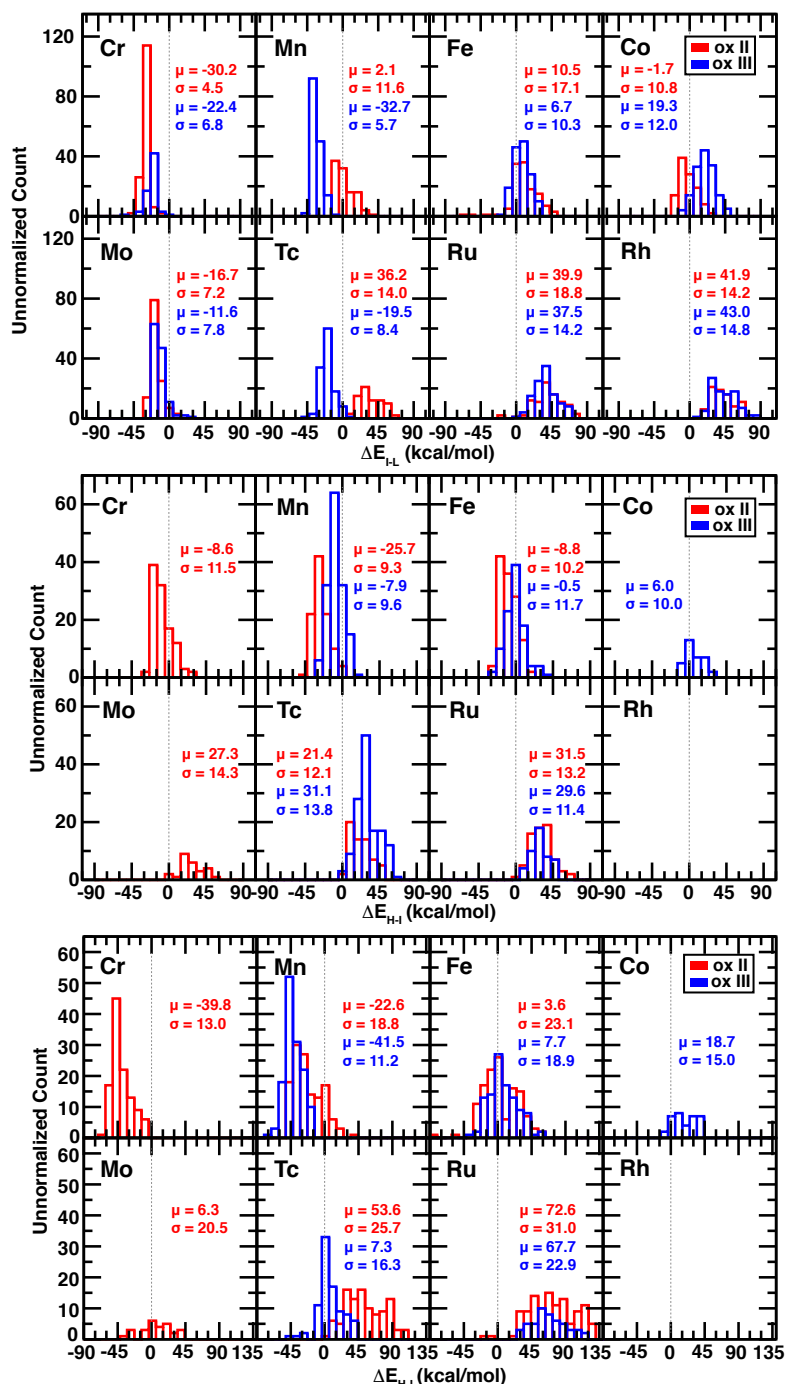

**Figure S6**. Distribution of adiabatic Δ*E* spin-splitting (in kcal/mol) for 3*d* TMCs (upper panes) and 4*d* TMCs (lower panes) in the full data set prior to the ILP or IMP pairing. Data is grouped by spin states and differences in number of paired electrons between states: 2-electron LS-IS (top), 2-electron IS-HS (middle), and 4-electron LS-HS (bottom). The zero value is shown as a vertical dashed line within each pane. Distributions are colored by oxidation state: II (red) and III (blue). Mean values (μ, in kcal/mol) and standard deviations (σ, in kcal/mol) of distributions are labeled within each pane by oxidation state. The majority of 4*d* TMCs favor LS states (positive values), whereas 3*d* TMC distributions have a higher population of higher-spin states (negative values).

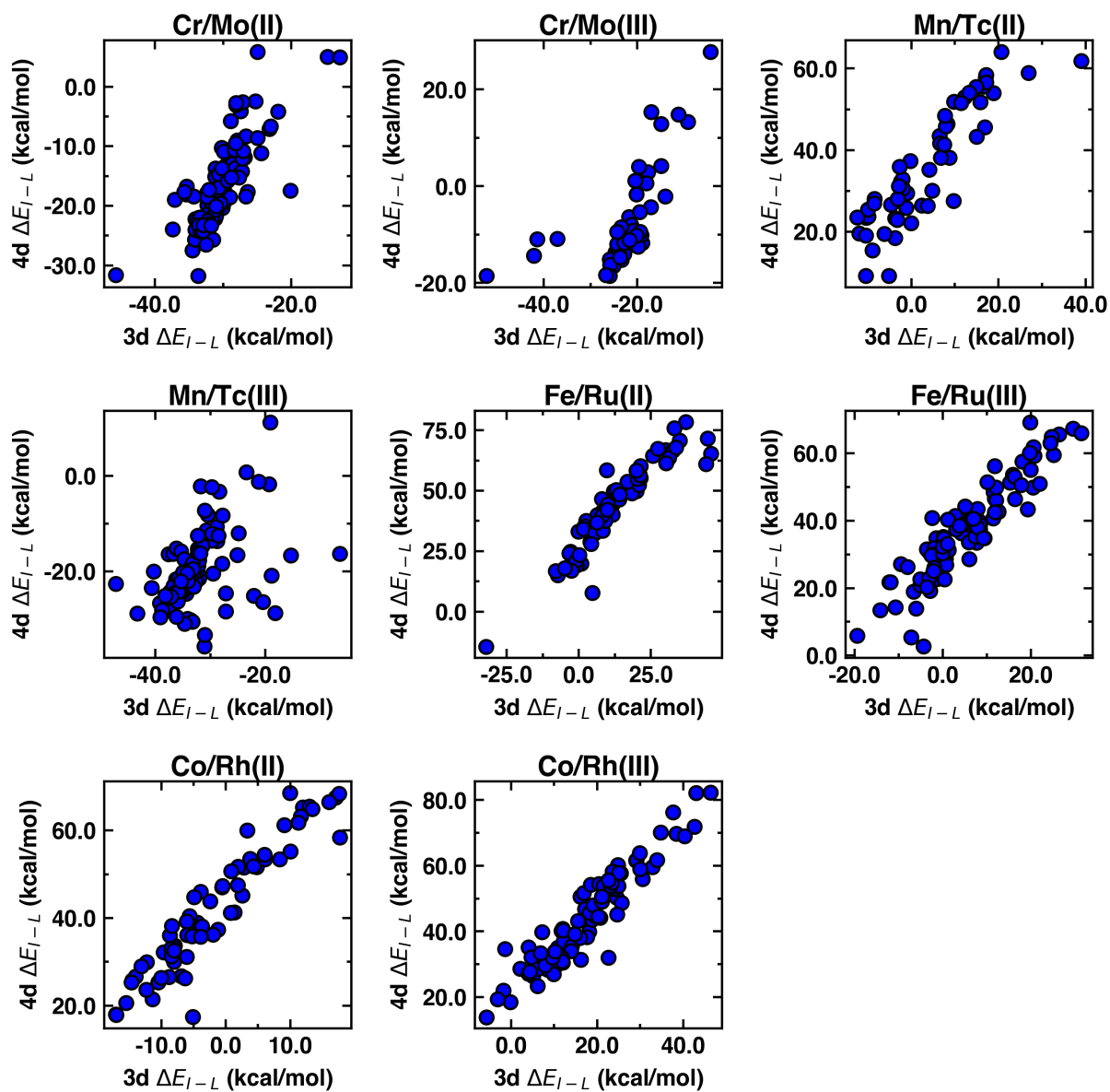

**Figure S7.** Parity plots of the 3d and 4d IS-LS spin-splitting, i.e., 3d $\Delta E_{I-L}$ and 4d $\Delta E_{I-L}$, (in kcal/mol) in the IMP data set. The range of the graphs is set to be the largest range in the data, which varies by metal and oxidation state and can be larger for one of the two axes.

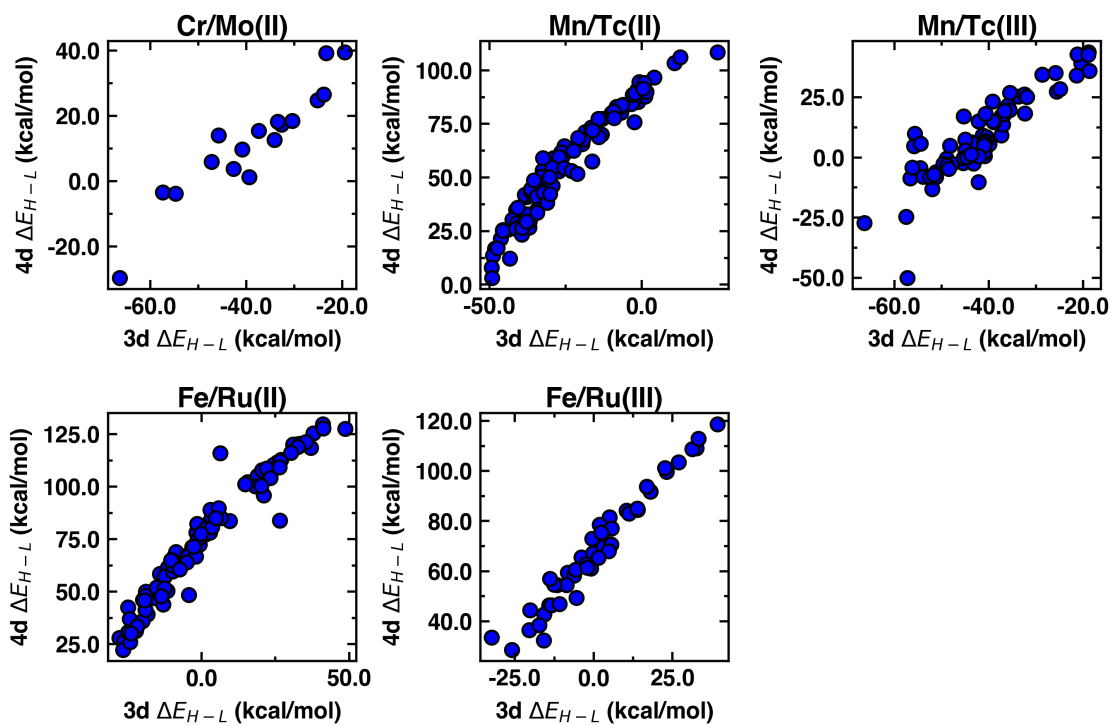

**Figure S8.** Parity plots of the *3d* and *4d* HS-LS spin-splitting, i.e., *3d* $\Delta E_{H\text{-}L}$ and *4d* $\Delta E_{H\text{-}L}$, (in kcal/mol) in the IMP data set. The range of the graphs is set to be the largest range in the data, which varies by metal and oxidation state and can be larger for one of the two axes.

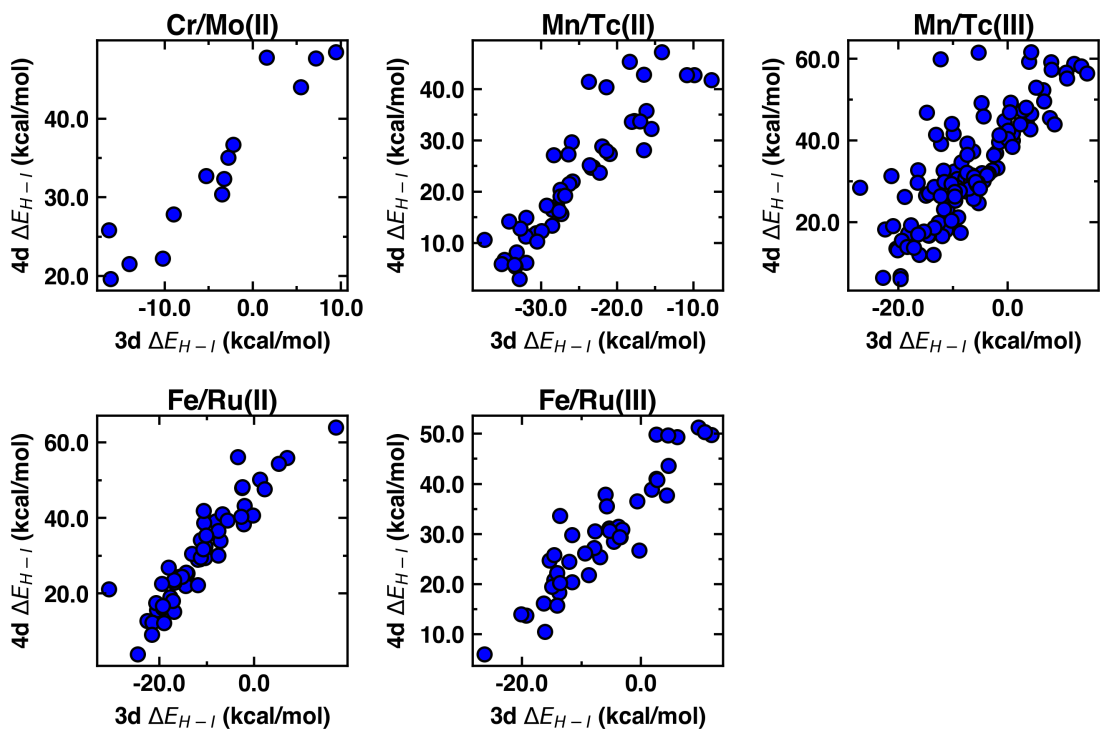

**Figure S9.** Parity plots of the *3d* and *4d* HS-IS spin-splitting, i.e., *3d* $\Delta E_{H\text{-}I}$ and *4d* $\Delta E_{H\text{-}I}$, (in kcal/mol) in the IMP data set. The range of the graphs is set to be the largest range in the data, which varies by metal and oxidation state and can be larger for one of the two axes.

**Table S11.** Linear models (y = ax + b, a, unitless, and b, in kcal/mol) relating 3*d* (x) and 4*d* (y) spin-splitting properties (specific spin states are annotated in table) for complexes in the IMP set by metal (M) and oxidation (ox) state. Models are only reported if at least five data points are available. The mean unsigned error (MUE) in predicting the 4*d* metal property from the 3*d* metal property using the linear model is reported along with the number of points in the set. Lines were fit using ordinary least squares, and the $R^2$ value for each line is shown.

| $\Delta E_{H-L}$ | | | | | |
|---|---|---|---|---|---|
| M(ox) | | Linear Model | | $R^2$ | MUE (kcal/mol) | # points |
| x | y | a | b | | | |
| Cr(II) | Mo(II) | 1.21 | 58.78 | 0.88 | 4.68 | 17 |
| Mn(III) | Tc(III) | 1.45 | 68.70 | 0.79 | 5.41 | 71 |
| Mn(II) | Tc(II) | 1.47 | 91.35 | 0.94 | 5.06 | 96 |
| Fe(III) | Ru(III) | 1.36 | 67.35 | 0.96 | 3.62 | 47 |
| Fe(II) | Ru(II) | 1.45 | 73.47 | 0.93 | 5.27 | 93 |
| Average MUE | | -- | -- | -- | 4.81 | -- |

| $\Delta E_{I-L}$ | | | | | |
|---|---|---|---|---|---|
| M(ox) | | Linear Model | | $R^2$ | MUE (kcal/mol) | # points |
| x | y | a | b | | | |
| Cr(III) | Mo(III) | 0.88 | 11.71 | 0.44 | 5.59 | 60 |
| Cr(II) | Mo(II) | 1.35 | 24.21 | 0.61 | 3.30 | 104 |
| Mn(III) | Tc(III) | 0.59 | -0.89 | 0.2 | 5.30 | 96 |
| Mn(II) | Tc(II) | 1.16 | 31.78 | 0.79 | 5.43 | 55 |
| Fe(III) | Ru(III) | 1.27 | 30.38 | 0.84 | 4.48 | 94 |
| Fe(II) | Ru(II) | 1.19 | 28.35 | 0.86 | 4.83 | 68 |
| Co(III) | Rh(III) | 1.27 | 21.81 | 0.88 | 4.05 | 86 |
| Co(II) | Rh(II) | 1.47 | 44.45 | 0.89 | 3.60 | 67 |
| Average MUE | | -- | -- | -- | 4.57 | -- |

| $\Delta E_{H-I}$ | | | | | |
|---|---|---|---|---|---|
| M(ox) | | Linear Model | | $R^2$ | MUE (kcal/mol) | # points |
| x | y | a | b | | | |
| Cr(II) | Mo(II) | 1.14 | 38.57 | 0.89 | 2.17 | 14 |
| Mn(III) | Tc(III) | 1.22 | 41.74 | 0.64 | 6.05 | 114 |
| Mn(II) | Tc(II) | 1.55 | 61.77 | 0.82 | 3.80 | 51 |
| Fe(III) | Ru(III) | 1.20 | 37.68 | 0.85 | 3.20 | 44 |
| Fe(II) | Ru(II) | 1.40 | 46.37 | 0.85 | 3.69 | 54 |
| Average MUE | | -- | -- | -- | 3.78 | -- |

**Table S12.** Frequency of metal and oxidation state combinations in the filtered data sets for 3$d$ and 4$d$ TMC adiabatic $\Delta E$ pairs grouped by 4-electron $\Delta E_{H-L}$, 2-electron $\Delta E_{I-L}$, and 2-electron $\Delta E_{H-I}$. When a spin state in the definition is undefined, no data was collected and -- is indicated, whereas when no data was converged and retained, a zero is indicated. Fewer late TMCs have converged $\Delta E$ values that involve high-spin states.

| 3$d$ | 4 e H-L | 2 e I-L | 2 e H-I | 4$d$ | 4 e H-L | 2 e I-L | 2 e H-I |
|---|---|---|---|---|---|---|---|
| **Full Set** | | | | | | | |
| Cr(III) | -- | 67 | -- | Mo(III) | -- | 126 | -- |
| Cr(II) | 110 | 149 | 106 | Mo(II) | 27 | 128 | 26 |
| Mn(III) | 136 | 158 | 148 | Tc(III) | 188 | 125 | 137 |
| Mn(II) | 129 | 105 | 101 | Tc(II) | 103 | 75 | 62 |
| Fe(III) | 107 | 153 | 101 | Ru(III) | 50 | 113 | 47 |
| Fe(II) | 127 | 123 | 125 | Ru(II) | 113 | 91 | 70 |
| Co(III) | 34 | 146 | 33 | Rh(III) | 0 | 96 | 0 |
| Co(II) | -- | 110 | -- | Rh(II) | -- | 71 | -- |
| **IMP Set** | | | | | | | |
| Cr(III) | -- | 60 | -- | Mo(III) | -- | 60 | -- |
| Cr(II) | 17 | 104 | 14 | Mo(II) | 17 | 104 | 14 |
| Mn(III) | 71 | 96 | 114 | Tc(III) | 71 | 96 | 114 |
| Mn(II) | 96 | 55 | 51 | Tc(II) | 96 | 55 | 51 |
| Fe(III) | 47 | 94 | 44 | Ru(III) | 47 | 94 | 44 |
| Fe(II) | 93 | 68 | 54 | Ru(II) | 93 | 68 | 54 |
| Co(III) | 0 | 86 | 0 | Rh(III) | 0 | 86 | 0 |
| Co(II) | -- | 67 | -- | Rh(II) | -- | 67 | -- |
| **ILP Set** | | | | | | | |
| Cr(III) | -- | 11 | -- | Mo(III) | -- | 30 | -- |
| Cr(II) | 21 | 26 | 26 | Mo(II) | 1 | 14 | 2 |
| Mn(III) | 27 | 34 | 36 | Tc(III) | 12 | 15 | 32 |
| Mn(II) | 31 | 14 | 16 | Tc(II) | 27 | 9 | 10 |
| Fe(III) | 8 | 28 | 7 | Ru(III) | 2 | 15 | 2 |
| Fe(II) | 26 | 21 | 25 | Ru(II) | 25 | 15 | 12 |
| Co(III) | 0 | 34 | 0 | Rh(III) | 0 | 12 | 0 |
| Co(II) | -- | 28 | -- | Rh(II) | -- | 15 | -- |

**Table S13.** Linear models (y = ax + b, a, unitless, and b, in kcal/mol) relating 3*d* (x) and 4*d* (y) $\Delta E_{LD}$ properties for complexes in the IMP set by metal (M), oxidation (ox) state, and spin multiplicity (2*S*+1). Models are only reported if at least five data points are available. The mean unsigned error (MUE) in predicting the 4*d* metal property from the 3*d* metal property using the linear model is reported along with the number of points in the set. Lines were fit using ordinary least squares, and the $R^2$ value for each line is shown.

| M(ox) | 2S+1 | Linear Model | | $R^2$ | MUE (kcal/mol) | # points |
|---|---|---|---|---|---|---|
| | | a | b | | | |
| Cr(III) | 2 | 0.90 | 3.31 | 0.88 | 8.95 | 23 |
| Cr(III) | 4 | 0.92 | 5.30 | 1.00 | 1.07 | 67 |
| Cr(II) | 1 | 0.88 | 9.91 | 0.85 | 3.10 | 44 |
| Cr(II) | 3 | 0.83 | 10.53 | 0.93 | 2.36 | 75 |
| Cr(II) | 5 | 0.89 | 0.22 | 0.41 | 7.79 | 13 |
| Mn(III) | 1 | 0.90 | 7.28 | 0.93 | 5.15 | 37 |
| Mn(III) | 3 | 0.95 | 5.22 | 0.98 | 2.44 | 74 |
| Mn(III) | 5 | 1.01 | -4.14 | 0.94 | 4.96 | 59 |
| Mn(II) | 2 | 0.90 | 10.37 | 0.92 | 2.44 | 69 |
| Mn(II) | 4 | 0.75 | 6.20 | 0.62 | 7.87 | 29 |
| Mn(II) | 6 | 0.92 | -1.71 | 0.97 | 1.33 | 55 |
| Fe(III) | 2 | 0.92 | 6.95 | 0.98 | 2.27 | 81 |
| Fe(III) | 4 | 0.96 | -1.10 | 0.93 | 4.84 | 34 |
| Fe(III) | 6 | 0.94 | -3.22 | 0.99 | 0.98 | 23 |
| Fe(II) | 1 | 0.80 | 11.38 | 0.84 | 2.39 | 68 |
| Fe(II) | 3 | 0.90 | 1.39 | 0.85 | 5.80 | 40 |
| Fe(II) | 5 | 0.88 | 0.27 | 0.97 | 1.56 | 51 |
| Co(III) | 1 | 0.80 | 12.50 | 0.88 | 3.39 | 92 |
| Co(III) | 3 | 0.95 | 1.96 | 0.90 | 5.88 | 38 |
| Co(II) | 2 | 0.88 | 3.68 | 0.74 | 6.42 | 38 |
| Co(II) | 4 | 0.79 | 3.23 | 0.94 | 1.69 | 47 |
| **Average** | -- | -- | -- | -- | 3.94 | -- |

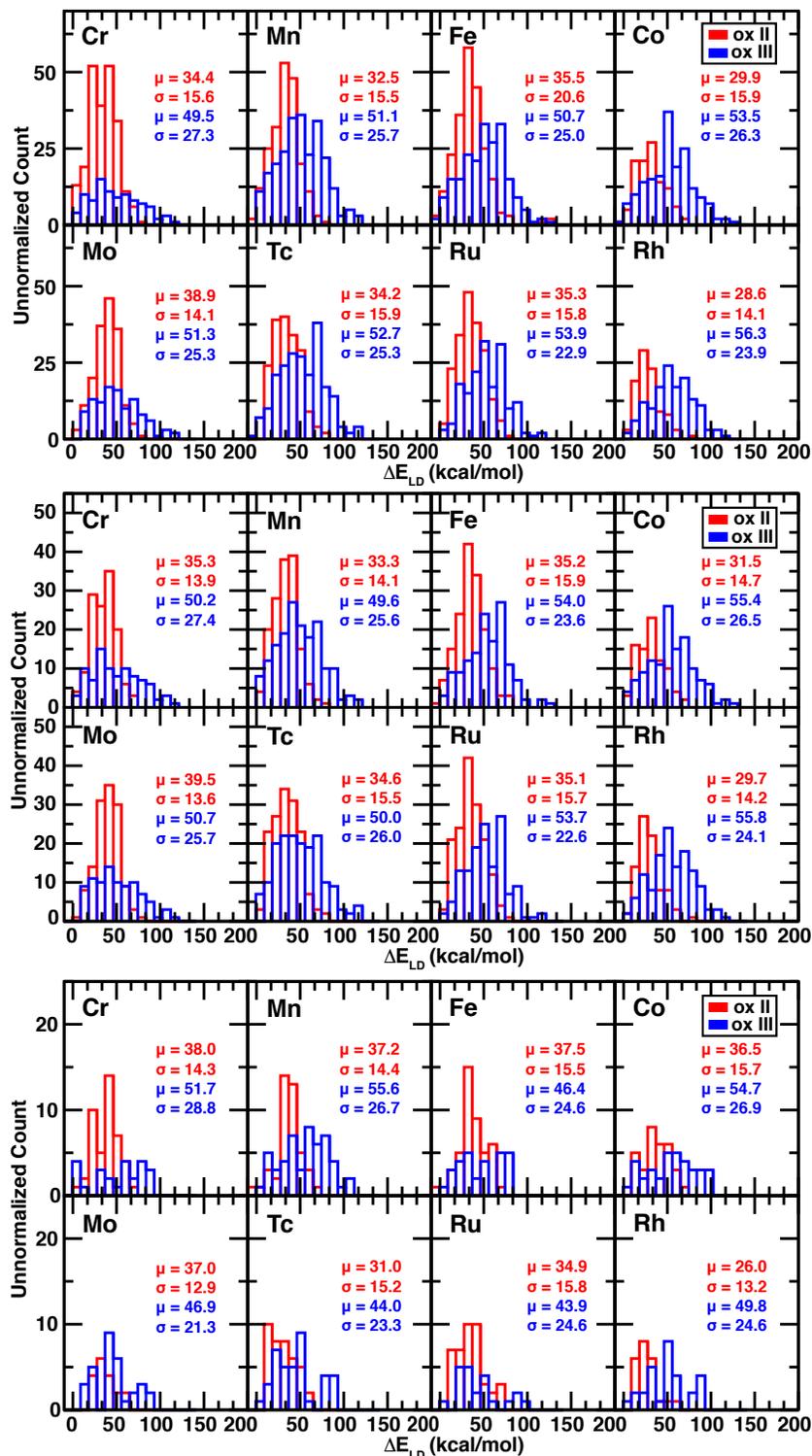

**Figure S10**. Distribution of $\Delta E_{LD}$ (in kcal/mol) for 3$d$ TMCs (upper panes) and 4$d$ TMCs (lower panes) in the full data set (top), IMP (middle), and ILP (bottom) data sets. Distributions are colored by oxidation state: II (red) and III (blue). Mean values ($\mu$, in kcal/mol) and standard deviations ($\sigma$, in kcal/mol) of distributions are labeled within each pane by oxidation state.

**Table S14**. Frequency of metal, oxidation, and spin-state combinations for the $\Delta E_{LD}$ property in the full data set, the IMP data set, and the ILP data set, as indicated in the table below. When a spin state in the definition is undefined, no data was collected and -- is indicated, whereas when no data was converged and retained, a zero is indicated.

| | Full Set | | | | | | |
|---|---|---|---|---|---|---|---|
| 3d | LD | | | 4d | LD | | |
| | LS | IS | HS | | LS | IS | HS |
| Cr(III) | 27 | 67 | -- | Mo(III) | 42 | 68 | -- |
| Cr(II) | 83 | 77 | 67 | Mo(II) | 73 | 82 | 15 |
| Mn(III) | 83 | 86 | 76 | Tc(III) | 59 | 92 | 67 |
| Mn(II) | 74 | 59 | 74 | Tc(II) | 86 | 47 | 55 |
| Fe(III) | 85 | 76 | 50 | Ru(III) | 93 | 54 | 23 |
| Fe(II) | 81 | 74 | 73 | Ru(II) | 89 | 47 | 62 |
| Co(III) | 94 | 73 | 11 | Rh(III) | 93 | 46 | 0 |
| Co(II) | 52 | 58 | -- | Rh(II) | 44 | 51 | -- |
| | IMP Set | | | | | | |
| Cr(III) | 23 | 67 | -- | Mo(III) | 23 | 67 | -- |
| Cr(II) | 44 | 75 | 13 | Mo(II) | 44 | 75 | 13 |
| Mn(III) | 37 | 74 | 59 | Tc(III) | 37 | 74 | 59 |
| Mn(II) | 69 | 29 | 55 | Tc(II) | 69 | 29 | 55 |
| Fe(III) | 81 | 34 | 23 | Ru(III) | 81 | 34 | 23 |
| Fe(II) | 68 | 40 | 51 | Ru(II) | 68 | 40 | 51 |
| Co(III) | 92 | 38 | 0 | Rh(III) | 92 | 38 | 0 |
| Co(II) | 38 | 47 | -- | Rh(II) | 38 | 47 | -- |
| | ILP Set | | | | | | |
| Cr(III) | 6 | 42 | -- | Mo(III) | 18 | 44 | -- |
| Cr(II) | 24 | 30 | 34 | Mo(II) | 14 | 20 | 2 |
| Mn(III) | 30 | 40 | 32 | Tc(III) | 16 | 32 | 30 |
| Mn(II) | 30 | 14 | 42 | Tc(II) | 32 | 14 | 32 |
| Fe(III) | 32 | 22 | 4 | Ru(III) | 38 | 6 | 0 |
| Fe(II) | 22 | 36 | 32 | Ru(II) | 38 | 22 | 24 |
| Co(III) | 33 | 26 | 0 | Rh(III) | 50 | 8 | 0 |
| Co(II) | 28 | 36 | -- | Rh(II) | 24 | 22 | -- |

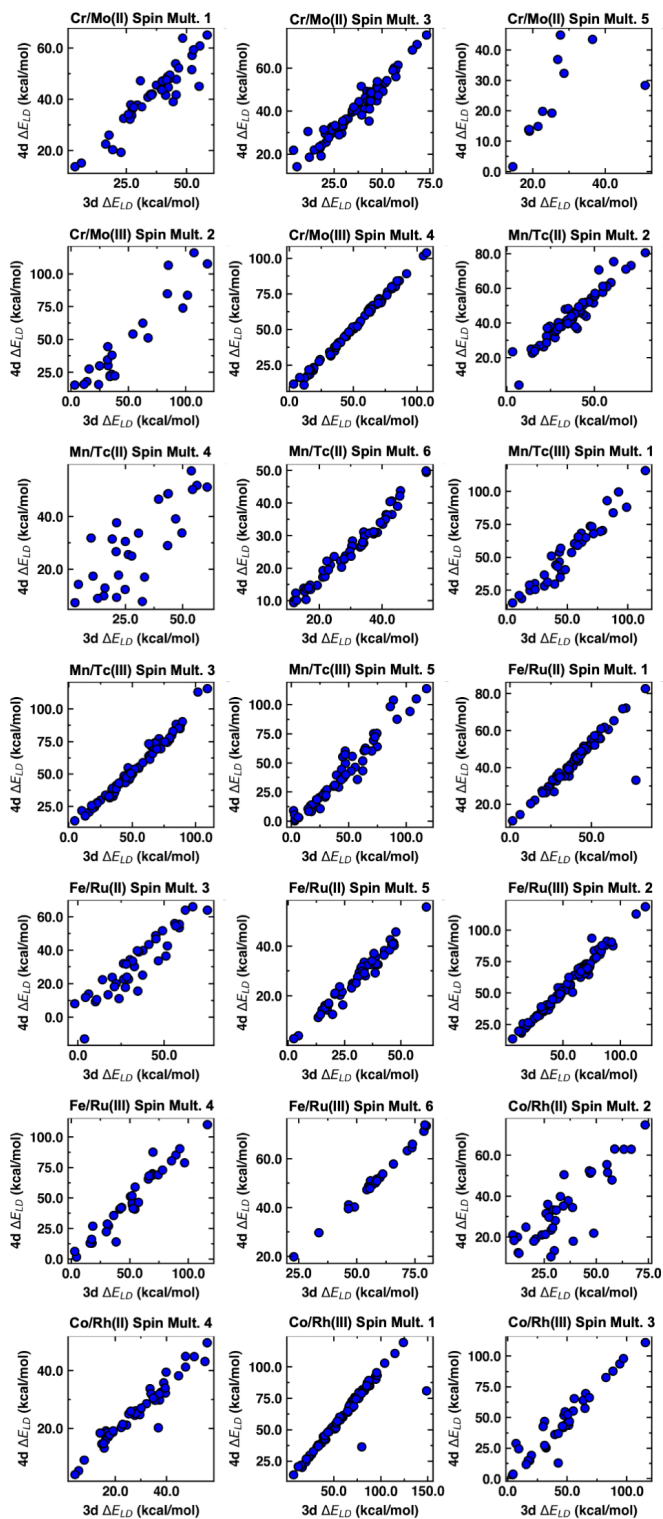

**Figure S11.** Parity plots of the 3*d* to 4*d* $\Delta E_{LD}$ (in kcal/mol) in the IMP data set. Sets are distinguished by metal/oxidation state and spin multiplicity (Spin. Mult.). The range of the graphs is set to be the largest range in the data, which varies and can be larger for one of the two axes.

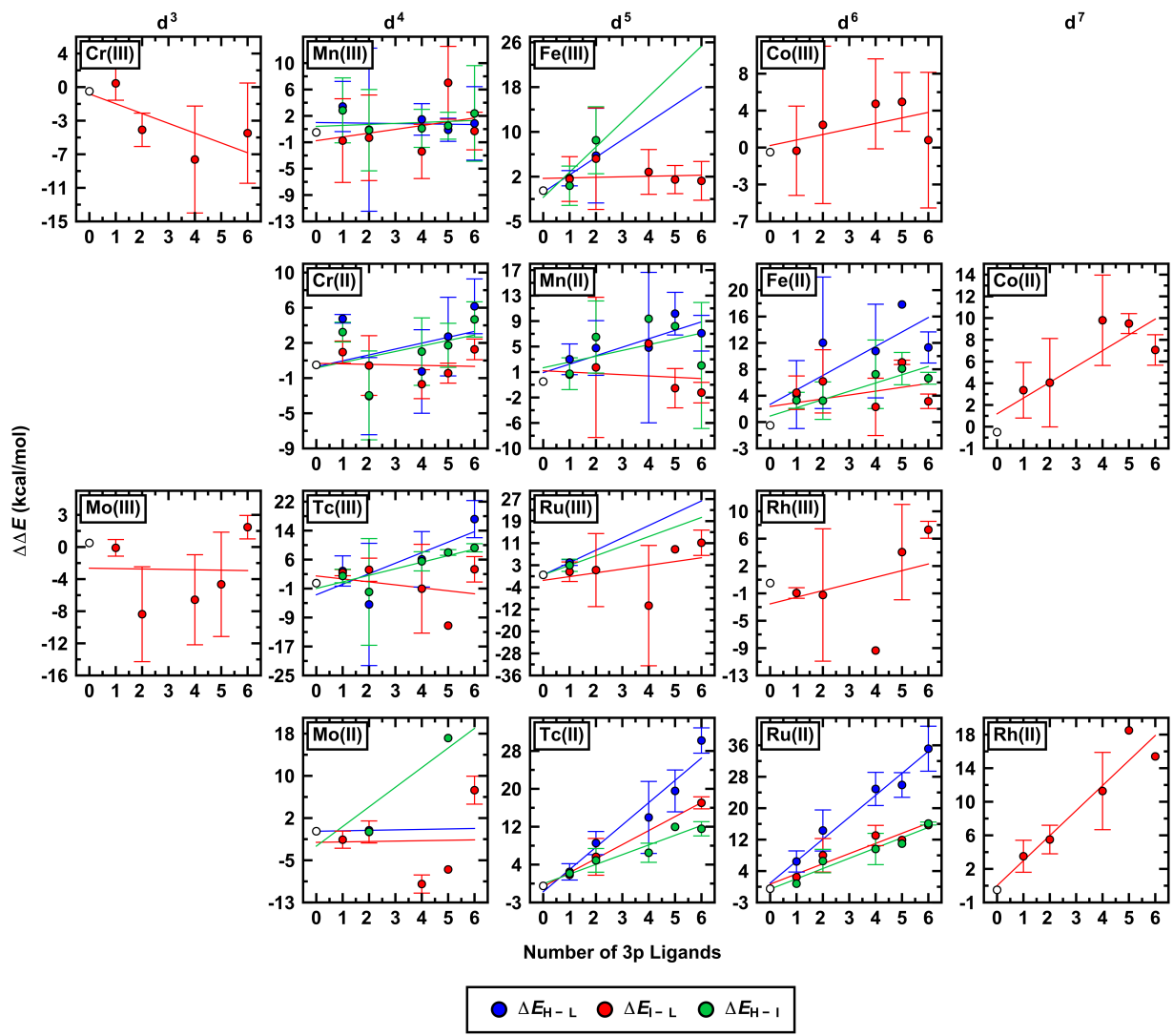

**Figure S12**. The change $\Delta\Delta E$ in spin-splitting energies, $\Delta E$, in kcal/mol in the ILP data set for H-L (blue circles), I-L (red circles), and H-I (green circles) spin state comparisons. The data is subdivided by metal and oxidation state and reported as a function of the number of ligands undergoing *2p* to *3p* substitution (i.e., 1, 2, 4, 5, or 6 ligands). We include an additional point shown as a white circle at (0,0) when fitting trend lines with ordinary least squares through average values. The error bars shown are standard deviations evaluated over all available pairs. The range of the y-axis varies with metal type, and the clearest trends are apparent for later transition metals.

**Table S15.** Linear fits of the change in spin-splitting energies (ΔΔ$E$, y in kcal/mol) in the ILP data set as a function of the number of ligands undergoing $2p$ to $3p$ substitution (x). Best-fit lines are fit using ordinary least squares, and an $R^2$ is reported for each fit.

|    | Oxidation State II | | | Oxidation State III | | |
|----|---|---|---|---|---|---|
|    | $\Delta E_{H-L}$ | $\Delta E_{I-L}$ | $\Delta E_{H-I}$ | $\Delta E_{H-L}$ | $\Delta E_{I-L}$ | $\Delta E_{H-I}$ |
| Cr | y= 0.669x -0.204 $R^2$=0.165 | y= -0.057x + 0.174 $R^2$=0.008 | y= 0.624x - 0.345 $R^2$=0.238 | | y= -1.168x - 0.315 $R^2$=0.558 | |
| Mn | y=1.328x + 1.41 $R^2$=0.756 | y=-0.205x + 1.698 $R^2$=0.03 | y=0.906x + 2.169 $R^2$=0.266 | y=-0.051x + 1.494 $R^2$=0.007 | y=0.574x - 1.262 $R^2$=0.14 | y=0.151x + 0.906 $R^2$=0.065 |
| Fe | y=2.196x + 3.177 $R^2$=0.647 | y=0.582x + 2.858 $R^2$=0.175 | y=1.251x + 1.417 $R^2$=0.828 | y=3.13x - 0.304 $R^2$=0.972 | y=0.096x + 2.185 $R^2$=0.014 | y=4.503x - 1.21 $R^2$=0.822 |
| Co | | y=1.454x + 1.682 $R^2$=0.744 | | | y=0.599x + 0.718 $R^2$=0.34 | |
| Mo | y=0.088x + 0.0 $R^2$=1.0 | y=0.078x - 2.101 $R^2$=0.001 | y=3.721x - 2.826 $R^2$=0.839 | | y=-0.049x - 3.141 $R^2$=0.001 | |
| Tc | y=4.704x - 1.239 $R^2$=0.952 | y=2.956x - 0.145 $R^2$=0.998 | y=2.051x + 0.446 $R^2$=0.942 | y=2.908x - 3.218 $R^2$=0.639 | y=-0.822x + 1.986 $R^2$=0.108 | y=1.811x - 1.471 $R^2$=0.774 |
| Ru | y=5.587x + 1.44 $R^2$=0.979 | y=2.589x + 1.168 $R^2$=0.931 | y=2.616x - 0.095 $R^2$=0.957 | y=4.469x + 0.0 $R^2$=1.0 | y=1.369x - 2.002 $R^2$=0.162 | y=3.47x + 0.0 $R^2$=1.0 |
| Rh | | y=2.981x + 0.504 $R^2$=0.925 | | | y=0.973x - 3.035 $R^2$=0.144 | |

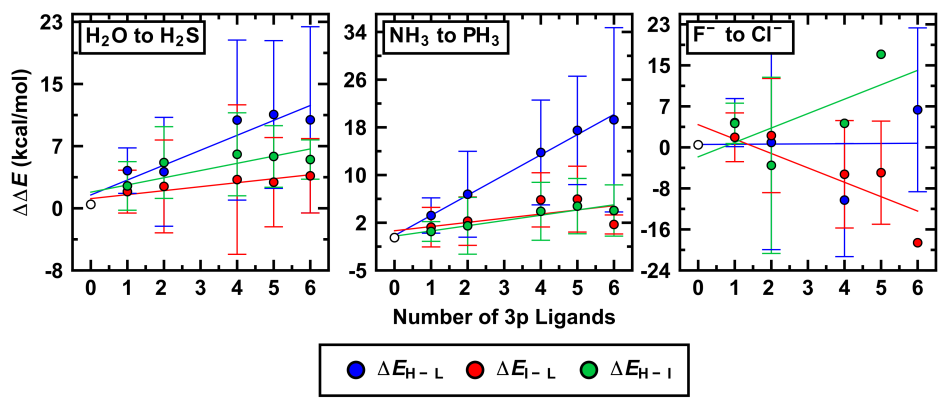

**Figure S13.** The change ΔΔ$E$ in spin-splitting energies, Δ$E$, in kcal/mol in the ILP data set as a function of the number of ligands undergoing $2p$ to $3p$ substitution (x, i.e., 1, 2, 4, 5, or 6 ligands), subdivided by the ligand undergoing substitution for H-L (blue circles), I-L (red circles), and H-I (green circles) spin state comparisons. The TMCs that include the simultaneous substitution of multiple ligand types are excluded. We include an additional point shown as a white circle at (0,0) when fitting trend lines with ordinary least squares through average values. The error bars shown are standard deviations evaluated over all available pairs.

**Table S16.** Linear fits of the change in spin-splitting energies (ΔΔ$E$, y in kcal/mol) in the ILP data set as a function of the number of ligands undergoing 2$p$ to 3$p$ substitution (x), subdivided by specific ligands undergoing substitution (indicated in left column). Best-fit lines are fit using ordinary least squares with a point at (0,0) included, and an $R^2$ is reported for each fit.

|  | Δ$E_{H-L}$ | Δ$E_{I-L}$ | Δ$E_{H-I}$ |
|---|---|---|---|
| $H_2O$ to $H_2S$ | y=1.92x+1.214 $R^2$=0.9 | y=0.513x+0.744 $R^2$=0.842 | y=0.927x+1.577 $R^2$=0.716 |
| $NH_3$ to $PH_3$ | y=3.388x+0.34 $R^2$=0.995 | y=0.7x+1.155 $R^2$=0.404 | y=0.878x+0.244 $R^2$=0.921 |
| $F^-$ to $Cl^-$ | y=0.039x+0.064 $R^2$=0.0 | y=-2.809x+3.926 $R^2$=0.71 | y=2.812x-2.344 $R^2$=0.511 |

**Table S17.** The average difference in Δ$E_{LD}$ in kcal/mol between TMCs with identical metals and axial ligands but with equatorial ligands that differ from 2$p$ to 3$p$ (i.e., $H_2O/S_2O$, $NH_3/PH_3$, or $F^-/Cl^-$). Positive values indicate that Δ$E_{LD}$ is larger in TMCs with equatorial 3$p$ ligands, and negative values indicate the opposite. Standard deviations for each subset are also shown.

| Cr(III) | Cr(II) | Mn(III) | Mn(II) | Fe(III) | Fe(II) | Co(III) | Co(II) |
|---|---|---|---|---|---|---|---|
| 4.82±7.88 | 6.67±9.31 | -0.14±8.92 | 4.04±2.92 | 2.69±6.11 | 4.33±3.46 | 1.54±6.57 | 5.74±7.38 |
| Mo(III) | Mo(II) | Tc(III) | Tc(II) | Ru(III) | Ru(II) | Rh(III) | Rh(II) |
| 0.62±2.2 | 1.71±5.6 | -2.4±8.79 | -1.99±8.53 | 2.8±2.6 | 3.48±3.6 | 1.09±4.65 | 3.74±5.42 |

**Table S18.** The average difference in Δ$E_{LD}$ in kcal/mol between TMCs with identical metals but either the two axial ligands (including the dissociated ligand) or the four equatorial ligands exchanged from 2$p$ to the isovalent 3$p$ (i.e., $H_2O/S_2O$, $NH_3/PH_3$, or $F^-/Cl^-$). Positive values indicate that Δ$E_{LD}$ is larger in TMCs with 3$p$ ligands, and negative values indicate the opposite. Standard deviations for each subset are also shown. No halides were computed because only neutral ligands were dissociated.

|  | Equatorial Substitution | Axial Substitution |
|---|---|---|
| $H_2O$ to $H_2S$ | 0.64±6.53 | -11.6±4.71 |
| $NH_3$ to $PH_3$ | 2.78±7.43 | -15.94±6.96 |
| $F^-$ to $Cl^-$ | 4.58±6.15 | no available data |

**Table S19.** Average mean unsigned error (MUE, in kcal/mol) for ensembles of 25 models assigned to specific inter-row learning tasks (far left column) with particular model types (second column from left). The number of inter-row data points corresponds to the number of examples in the training set which come from the same row as the validation data.

|  |  | Inter-row Data Points | | | | | |
|---|---|---|---|---|---|---|---|
|  |  | 10 | 15 | 20 | 30 | 40 | 50 |
| 3$d$ to 4$d$ IMP $\Delta E_{H-L}$ | S-RACs | 18.3 | 14.7 | 12.7 | 10.9 | 9.4 | 7.9 |
|  | S-eRACs | 16.7 | 13.5 | 12.0 | 10.2 | 9.2 | 8.5 |
|  | T-RACs | 13.7 | 12.4 | 10.8 | 9.9 | 8.8 | 8.3 |
|  | T-eRACs | 11.0 | 10.0 | 9.1 | 8.4 | 7.8 | 7.4 |
| 2$p$ to 3$p$ ILP $\Delta E_{H-L}$ | S-RACs | 16.9 | 12.6 | 11.3 | 9.8 | 8.8 | 7.1 |
|  | S-eRACs | 15.8 | 14.6 | 13.0 | 11.1 | 10.1 | 9.4 |
|  | T-RACs | 8.4 | 7.3 | 6.4 | 5.9 | 5.4 | 5.0 |
|  | T-eRACs | 6.1 | 5.7 | 5.0 | 4.7 | 4.6 | 4.3 |
| 3$d$ to 4$d$ IMP $\Delta E_{LD}$ | S-RACs | 15.6 | 14.7 | 13.3 | 11.6 | 10.4 | 9.6 |
|  | S-eRACs | 15.8 | 14.6 | 13.0 | 11.1 | 10.1 | 9.4 |
|  | T-RACs | 7.5 | 7.4 | 7.2 | 6.9 | 6.7 | 6.5 |
|  | T-eRACs | 6.6 | 6.6 | 6.4 | 6.2 | 6.1 | 5.9 |
| 2$p$ to 3$p$ ILP $\Delta E_{LD}$ | S-RACs | 13.1 | 12.3 | 11.3 | 10.0 | 9.0 | 8.1 |
|  | S-eRACs | 13.4 | 12.4 | 11.2 | 9.8 | 8.6 | 7.9 |
|  | T-RACs | 7.9 | 7.6 | 7.1 | 6.7 | 6.5 | 6.2 |
|  | T-eRACs | 7.1 | 7.0 | 6.8 | 6.4 | 6.2 | 6.0 |

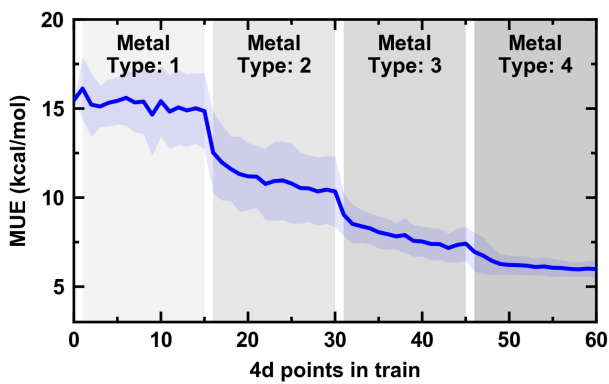

**Figure S14.** Mean unsigned error (MUE, in kcal/mol) results for the *3d* to *4d* learning task in the IMP data set on the $\Delta E_{I-L}$ property when the *4d* metals are grouped by type with a T-eRAC KRR model. The blue translucent shaded region represents the standard deviation of an ensemble of 25 models. For each model, the *4d* metal types (Mo, Tc, Ru, Rh) are arranged in a random order, and 15 data points corresponding to each metal are added before adding the next metal, as indicated by the gray shaded regions and white separators.

**Table S20.** Average mean unsigned errors (MUE, in kcal/mol) from an ensemble of 25 models for the typed inter-row learning task with a T-eRAC KRR model. Values are obtained for predicting the $\Delta E_{I-L}$ property in the IMP data set for the *3d* to *4d* learning task. In typed inter-row learning, the *4d* metals (Mo, Tc, Ru, Rh) are arranged in a random order, and 15 data points corresponding to each metal are added before moving to the next metal. We report the average MUE across the first 5 examples of each metal, the middle 5 examples, and the last 5 examples. We also report the average of all 15 MUEs for each metal added.

|  | First 5 | Middle 5 | Last 5 | Average of all 15 MUEs |
|---|---|---|---|---|
| Metal 1 | 15.4 | 15.3 | 14.9 | 15.2 |
| Metal 2 | 11.7 | 10.9 | 10.4 | 11.0 |
| Metal 3 | 8.5 | 7.8 | 7.3 | 7.8 |
| Metal 4 | 6.5 | 6.1 | 6.0 | 6.2 |

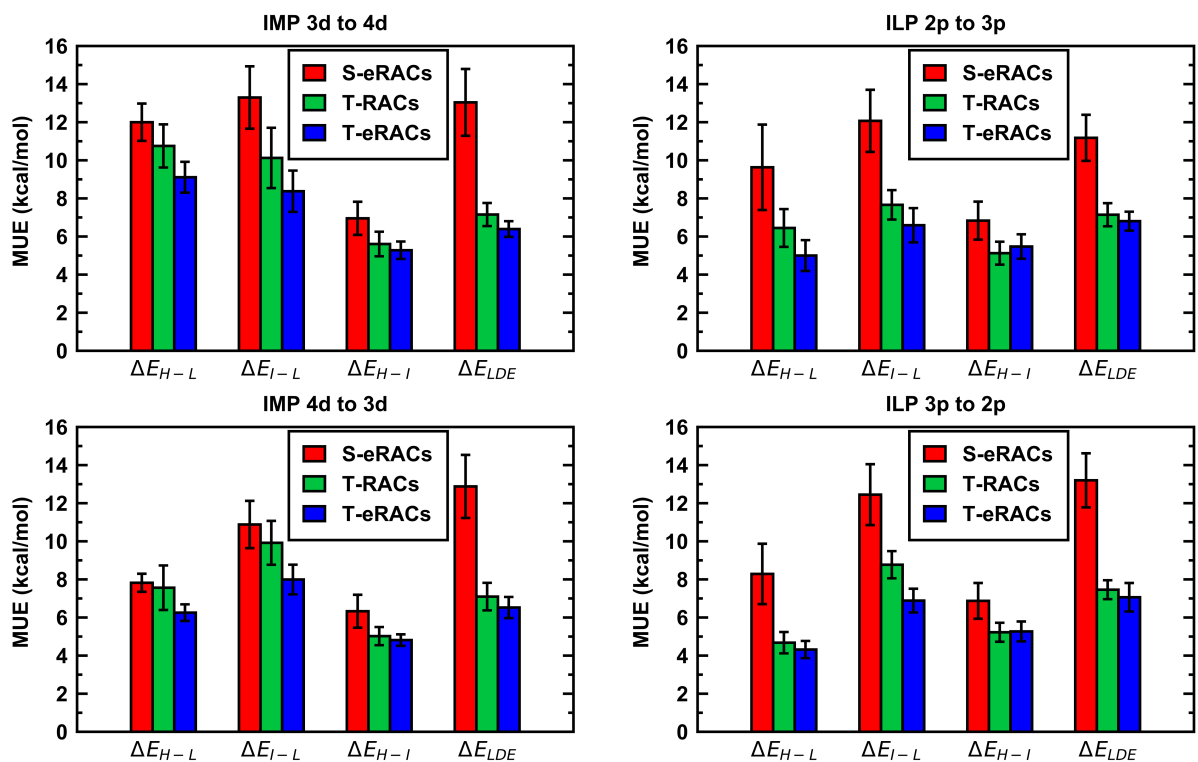

**Figure S15.** Mean unsigned error (MUE, in kcal/mol) results in a grouped bar graph by property (i.e., $\Delta E_{\text{H-L}}$, $\Delta E_{\text{I-L}}$, $\Delta E_{\text{H-I}}$, or $\Delta E_{\text{LD}}$) for each of the four learning tasks: learning $4d$ properties based on $3d$ data in the IMP data set (top left), learning $3p$ properties based on $2p$ data in the ILP data set (top right), learning $3d$ properties based on $4d$ data in the IMP data set (bottom left), and learning $2p$ properties based on $3p$ data in ILP data set (bottom right). The average of model performance is shown as the bar, the standard deviation over the ensemble is shown as an error bar, and the nature of the model (i.e., S-eRACs, T-RACs, or T-eRACs) is colored according to inset legend. In each case, the training set includes all available data from the alternate row of the periodic table, and the T models are also trained on 20 examples from the same row as the validation data.

**Table S21.** Mean unsigned error (MUE, in kcal/mol) for KRR model predictions with the average MUE of a 25-model ensemble shown along with its standard deviation. Results are reported for each of the four learning tasks, with four properties and three model types in each case. In each case, the training set includes all available data from the alternate row of the periodic table, and the T models are also trained on 20 examples from the same row as the validation data.

|  | ΔE$_{H-L}$ | ΔE$_{I-L}$ | ΔE$_{H-I}$ | ΔE$_{LD}$ |
|---|---|---|---|---|
| IMP 3*d* to 4*d* | | | | |
| S-eRACs | 12.0±0.98 | 13.3±1.64 | 6.95±0.87 | 13.04±1.75 |
| T-RACs | 10.75±1.13 | 10.12±1.58 | 5.61±0.65 | 7.15±0.61 |
| T-eRACs | 9.11±0.81 | 8.38±1.08 | 5.28±0.46 | 6.39±0.41 |
| ILP 2*p* to 3*p* | | | | |
| S-eRACs | 9.63±2.24 | 12.07±1.63 | 6.83±1.0 | 11.18±1.21 |
| T-RACs | 6.45±0.99 | 7.66±0.77 | 5.13±0.6 | 7.14±0.61 |
| T-eRACs | 5.0±0.81 | 6.59±0.9 | 5.47±0.64 | 6.81±0.5 |
| IMP 4*d* to 3*d* | | | | |
| S-eRACs | 7.82±0.47 | 10.88±1.24 | 6.33±0.86 | 12.88±1.65 |
| T-RACs | 7.56±1.17 | 9.92±1.15 | 5.02±0.48 | 7.1±0.72 |
| T-eRACs | 6.25±0.44 | 7.99±0.78 | 4.81±0.3 | 6.52±0.55 |
| ILP 3*p* to 2*p* | | | | |
| S-eRACs | 8.29±1.58 | 12.45±1.6 | 6.87±0.94 | 13.2±1.42 |
| T-RACs | 4.68±0.56 | 8.77±0.71 | 5.23±0.5 | 7.46±0.5 |
| T-eRACs | 4.32±0.45 | 6.89±0.62 | 5.27±0.52 | 7.06±0.75 |

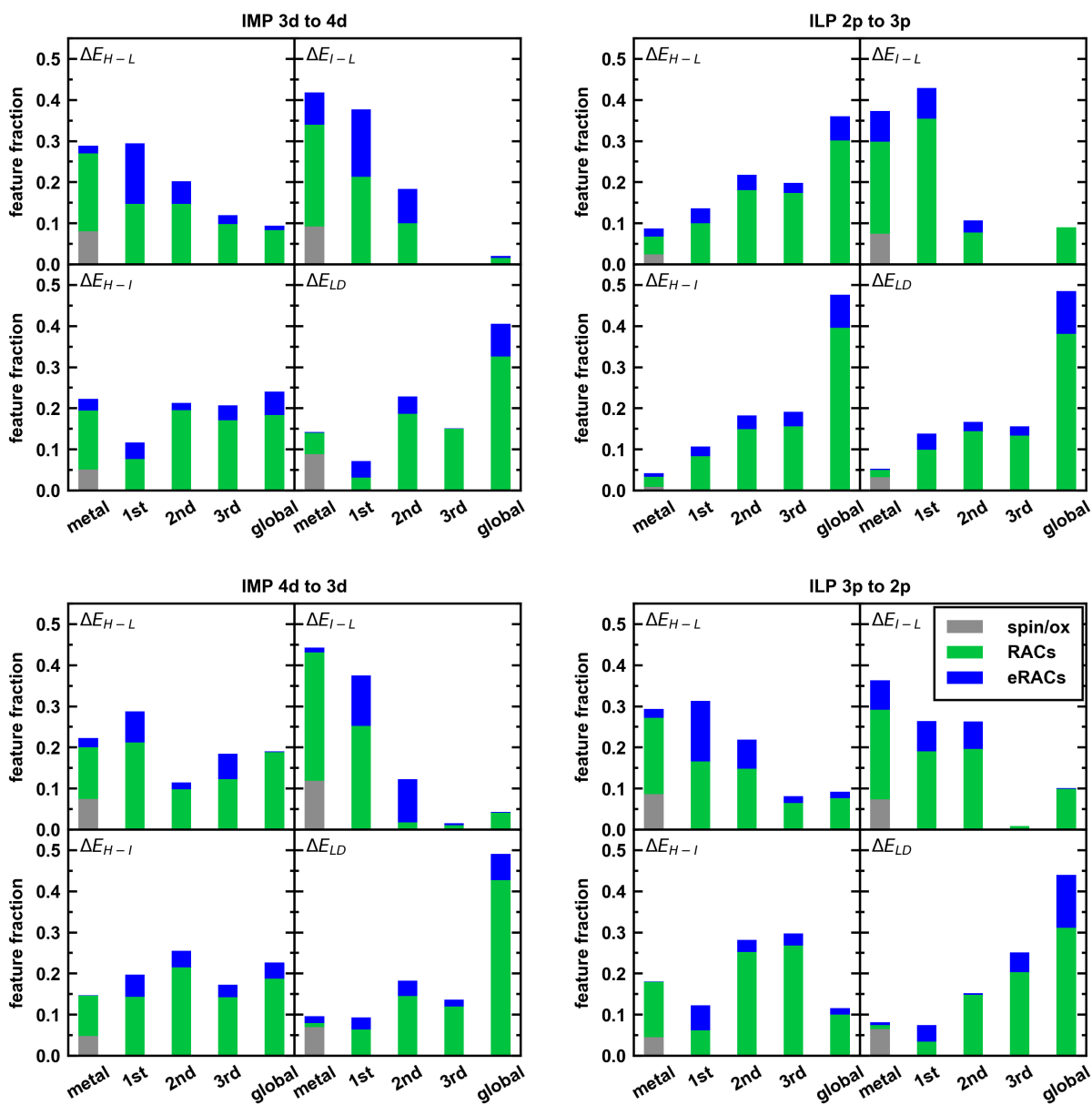

**Figure S16.** Fraction of features selected as a stacked bar chart in both the IMP (top) and ILP (bottom) data set when training models for all properties considered in this work (i.e., $\Delta E_{H\text{-}L}$, $\Delta E_{I\text{-}L}$, $\Delta E_{H\text{-}I}$, $\Delta E_{LD}$). In each case, 25 models are trained using 20 training points from the same row as the validation data and all data from the opposite row (i.e., T-eRAC models). Features are grouped by the atom that is most distant from the metal (0, 1, 2, 3, or > 3 bonds away) in the feature. Features are colored based on whether they are not based on autocorrelations (gray, spin/ox), are standard autocorrelation features in RAC-155 (green, RACs), or are derived from the $Z_{eff}$ atomic property (blue, eRACs).

Page S29

**Table S22.** Average fraction of features by heuristic property type in the feature-selected eRAC-185 subspace of the 25 KRR models used to build the ensemble for each property and data set indicated in the left two columns of the table below. Feature fractions are normalized prior to averaging across the 25 models, and any models that used the full representation (i.e., in lieu of LASSO or 1% RF cutoff) are excluded.

| | | ox/spin | $Z_{eff}$ | chi | Z | T | I | S |
|---|---|---|---|---|---|---|---|---|
| IMP 3d to 4d | $\Delta E_{H-L}$ | 0.08 | 0.25 | 0.26 | 0.32 | 0.02 | 0.01 | 0.07 |
| | $\Delta E_{I-L}$ | 0.09 | 0.33 | 0.20 | 0.27 | 0.00 | 0.00 | 0.10 |
| | $\Delta E_{H-I}$ | 0.05 | 0.18 | 0.21 | 0.33 | 0.02 | 0.05 | 0.16 |
| | $\Delta E_{LD}$ | 0.09 | 0.17 | 0.27 | 0.16 | 0.04 | 0.05 | 0.24 |
| ILP 2p to 3p | $\Delta E_{H-L}$ | 0.02 | 0.18 | 0.22 | 0.22 | 0.12 | 0.07 | 0.17 |
| | $\Delta E_{I-L}$ | 0.07 | 0.18 | 0.21 | 0.34 | 0.00 | 0.00 | 0.20 |
| | $\Delta E_{H-I}$ | 0.01 | 0.18 | 0.18 | 0.19 | 0.15 | 0.11 | 0.18 |
| | $\Delta E_{LD}$ | 0.03 | 0.19 | 0.19 | 0.17 | 0.12 | 0.10 | 0.19 |
| IMP 4d to 3d | $\Delta E_{H-L}$ | 0.08 | 0.18 | 0.26 | 0.36 | 0.01 | 0.02 | 0.11 |
| | $\Delta E_{I-L}$ | 0.12 | 0.25 | 0.21 | 0.24 | 0.00 | 0.00 | 0.18 |
| | $\Delta E_{H-I}$ | 0.05 | 0.16 | 0.26 | 0.35 | 0.04 | 0.05 | 0.10 |
| | $\Delta E_{LD}$ | 0.07 | 0.16 | 0.32 | 0.13 | 0.09 | 0.07 | 0.15 |
| ILP 3p to 2p | $\Delta E_{H-L}$ | 0.09 | 0.27 | 0.26 | 0.3 | 0.03 | 0.01 | 0.04 |
| | $\Delta E_{I-L}$ | 0.07 | 0.21 | 0.26 | 0.31 | 0.00 | 0.00 | 0.14 |
| | $\Delta E_{H-I}$ | 0.05 | 0.14 | 0.28 | 0.27 | 0.04 | 0.02 | 0.21 |
| | $\Delta E_{LD}$ | 0.06 | 0.23 | 0.18 | 0.22 | 0.06 | 0.07 | 0.18 |

**Table S23.** Average fraction of features categorized by the shortest bond path between the metal center and the most distant atom in the feature (i.e., 0, 1, 2, 3, or more and labeled as metal, 1st, 2nd, 3rd, or global) in the feature-selected eRAC-185 subspace of the 25 KRR models used to build the ensemble for each property and data set indicated in the left two columns of the table below. Feature fractions are normalized prior to averaging across the 25 models, and any models that used the full representation (i.e., in lieu of LASSO or 1% RF cutoff) are excluded.

|  |  | metal | 1st | 2nd | 3rd | global |
|---|---|---|---|---|---|---|
| IMP 3$d$ to 4$d$ | $\Delta E_{H-L}$ | 0.29 | 0.29 | 0.20 | 0.12 | 0.09 |
|  | $\Delta E_{I-L}$ | 0.42 | 0.38 | 0.18 | 0.00 | 0.02 |
|  | $\Delta E_{H-I}$ | 0.22 | 0.12 | 0.21 | 0.21 | 0.24 |
|  | $\Delta E_{LD}$ | 0.14 | 0.07 | 0.23 | 0.15 | 0.41 |
| ILP 2$p$ to 3$p$ | $\Delta E_{H-L}$ | 0.09 | 0.14 | 0.22 | 0.20 | 0.36 |
|  | $\Delta E_{I-L}$ | 0.37 | 0.43 | 0.11 | 0.00 | 0.09 |
|  | $\Delta E_{H-I}$ | 0.04 | 0.11 | 0.18 | 0.19 | 0.48 |
|  | $\Delta E_{LD}$ | 0.05 | 0.14 | 0.17 | 0.16 | 0.49 |
| IMP 4$d$ to 3$d$ | $\Delta E_{H-L}$ | 0.22 | 0.29 | 0.12 | 0.18 | 0.19 |
|  | $\Delta E_{I-L}$ | 0.44 | 0.38 | 0.12 | 0.02 | 0.04 |
|  | $\Delta E_{H-I}$ | 0.15 | 0.20 | 0.26 | 0.17 | 0.23 |
|  | $\Delta E_{LD}$ | 0.10 | 0.09 | 0.18 | 0.14 | 0.49 |
| ILP 3$p$ to 2$p$ | $\Delta E_{H-L}$ | 0.29 | 0.31 | 0.22 | 0.08 | 0.09 |
|  | $\Delta E_{I-L}$ | 0.36 | 0.26 | 0.26 | 0.01 | 0.10 |
|  | $\Delta E_{H-I}$ | 0.18 | 0.12 | 0.28 | 0.30 | 0.12 |
|  | $\Delta E_{LD}$ | 0.08 | 0.07 | 0.15 | 0.25 | 0.44 |

**Table S24.** Average distance between TMCs in the eRAC-185 and RAC-155 feature-selected subspaces of trained models for 3*d* to 4*d* in the IMP data set. TMCs are subdivided by the metal identity. The averages are weighted such that each of the 25 KRR models contributes equally, and distances are normalized such that the average distance between any two TMCs is 1.

| | | RAC-155 | | | | | | | | eRAC-185 | | | | | | | |
|---|---|---|---|---|---|---|---|---|---|---|---|---|---|---|---|---|---|
| | | Cr | Mn | Fe | Co | Mo | Tc | Ru | Rh | Cr | Mn | Fe | Co | Mo | Tc | Ru | Rh |
| ΔE_{H-L} | Cr | 0.86 | 0.93 | 0.90 | | 1.05 | 1.12 | 1.08 | | 0.60 | 0.82 | 1.01 | | 1.05 | 1.16 | 1.32 | |
| | Mn | 0.93 | 0.89 | 0.92 | | 1.1 | 1.08 | 1.1 | | 0.82 | 0.74 | 0.89 | | 1.18 | 1.13 | 1.26 | |
| | Fe | 0.90 | 0.92 | 0.87 | | 1.02 | 1.06 | 1.01 | | 1.01 | 0.89 | 0.74 | | 1.22 | 1.12 | 1.07 | |
| | Co | | | | | | | | | | | | | | | | |
| | Mo | 1.05 | 1.10 | 1.02 | | 0.89 | 0.98 | 0.94 | | 1.05 | 1.18 | 1.22 | | 0.65 | 0.87 | 1.05 | |
| | Tc | 1.12 | 1.08 | 1.06 | | 0.98 | 0.93 | 1.00 | | 1.16 | 1.13 | 1.12 | | 0.87 | 0.79 | 0.94 | |
| | Ru | 1.08 | 1.10 | 1.01 | | 0.94 | 1.00 | 0.91 | | 1.32 | 1.26 | 1.07 | | 1.05 | 0.94 | 0.79 | |
| | Rh | | | | | | | | | | | | | | | | |
| ΔE_{I-L} | Cr | 0.58 | 0.71 | 0.64 | 0.67 | 1.22 | 1.41 | 1.30 | 1.39 | 0.52 | 0.70 | 0.85 | 1.09 | 1.03 | 1.24 | 1.26 | 1.48 |
| | Mn | 0.71 | 0.61 | 0.78 | 0.81 | 1.26 | 1.31 | 1.39 | 1.49 | 0.70 | 0.57 | 0.79 | 0.97 | 1.11 | 1.12 | 1.24 | 1.42 |
| | Fe | 0.64 | 0.78 | 0.56 | 0.58 | 1.09 | 1.31 | 1.12 | 1.21 | 0.85 | 0.79 | 0.58 | 0.68 | 1.10 | 1.18 | 0.99 | 1.10 |
| | Co | 0.67 | 0.81 | 0.58 | 0.57 | 1.03 | 1.26 | 1.07 | 1.16 | 1.09 | 0.97 | 0.68 | 0.59 | 1.27 | 1.27 | 1.01 | 1.03 |
| | Mo | 1.22 | 1.26 | 1.09 | 1.03 | 0.67 | 0.87 | 0.78 | 0.81 | 1.03 | 1.11 | 1.10 | 1.27 | 0.57 | 0.81 | 0.91 | 1.15 |
| | Tc | 1.41 | 1.31 | 1.31 | 1.26 | 0.87 | 0.68 | 1.09 | 1.12 | 1.24 | 1.12 | 1.18 | 1.27 | 0.81 | 0.62 | 1.00 | 1.16 |
| | Ru | 1.30 | 1.39 | 1.12 | 1.07 | 0.78 | 1.09 | 0.67 | 0.70 | 1.26 | 1.24 | 0.99 | 1.01 | 0.91 | 1.00 | 0.63 | 0.73 |
| | Rh | 1.39 | 1.49 | 1.21 | 1.16 | 0.81 | 1.12 | 0.70 | 0.69 | 1.48 | 1.42 | 1.10 | 1.03 | 1.15 | 1.16 | 0.73 | 0.65 |
| ΔE_{H-I} | Cr | 0.81 | 0.92 | 0.89 | | 1.03 | 1.12 | 1.09 | | 0.75 | 0.90 | 0.98 | | 1.01 | 1.13 | 1.18 | |
| | Mn | 0.92 | 0.87 | 0.92 | | 1.1 | 1.08 | 1.11 | | 0.90 | 0.83 | 0.93 | | 1.11 | 1.07 | 1.15 | |
| | Fe | 0.89 | 0.92 | 0.86 | | 1.03 | 1.07 | 1.02 | | 0.98 | 0.93 | 0.83 | | 1.12 | 1.1 | 1.02 | |
| | Co | | | | | | | | | | | | | | | | |
| | Mo | 1.03 | 1.1 | 1.03 | | 0.84 | 0.96 | 0.93 | | 1.01 | 1.11 | 1.12 | | 0.78 | 0.95 | 1.02 | |
| | Tc | 1.12 | 1.08 | 1.07 | | 0.96 | 0.91 | 0.99 | | 1.13 | 1.07 | 1.10 | | 0.95 | 0.86 | 1.01 | |
| | Ru | 1.09 | 1.11 | 1.02 | | 0.93 | 0.99 | 0.90 | | 1.18 | 1.15 | 1.02 | | 1.02 | 1.01 | 0.87 | |
| | Rh | | | | | | | | | | | | | | | | |
| ΔE_{LD} | Cr | 0.95 | 0.96 | 0.95 | 0.95 | 1.03 | 1.05 | 1.04 | 1.05 | 0.93 | 0.96 | 0.97 | 1.01 | 1.00 | 1.05 | 1.04 | 1.10 |
| | Mn | 0.96 | 0.95 | 0.97 | 0.97 | 1.04 | 1.04 | 1.05 | 1.07 | 0.96 | 0.93 | 0.97 | 1.00 | 1.03 | 1.03 | 1.05 | 1.10 |
| | Fe | 0.95 | 0.97 | 0.94 | 0.94 | 1.01 | 1.03 | 1.00 | 1.02 | 0.97 | 0.97 | 0.92 | 0.94 | 1.01 | 1.03 | 0.97 | 1.00 |
| | Co | 0.95 | 0.97 | 0.94 | 0.93 | 1.00 | 1.02 | 1.00 | 1.00 | 1.01 | 1.00 | 0.94 | 0.92 | 1.04 | 1.05 | 0.97 | 0.97 |
| | Mo | 1.03 | 1.04 | 1.01 | 1.00 | 0.97 | 0.99 | 0.98 | 0.98 | 1.00 | 1.03 | 1.01 | 1.04 | 0.93 | 0.98 | 0.98 | 1.02 |
| | Tc | 1.05 | 1.04 | 1.03 | 1.02 | 0.99 | 0.97 | 1.01 | 1.01 | 1.05 | 1.03 | 1.03 | 1.05 | 0.98 | 0.94 | 1.01 | 1.05 |
| | Ru | 1.04 | 1.05 | 1.00 | 1.00 | 0.98 | 1.01 | 0.96 | 0.96 | 1.04 | 1.05 | 0.97 | 0.97 | 0.98 | 1.01 | 0.93 | 0.94 |
| | Rh | 1.05 | 1.07 | 1.02 | 1.00 | 0.98 | 1.01 | 0.96 | 0.96 | 1.10 | 1.10 | 1.00 | 0.97 | 1.02 | 1.05 | 0.94 | 0.93 |